\documentclass[12pt]{article}

\usepackage{scicite}
\usepackage{graphicx}% Include figure files
\usepackage{amsmath}
\usepackage{inputenc}

\usepackage{times}

\newenvironment{sciabstract}{%
\begin{quote} \bf}
{\end{quote}}

\title{Superadiabatic population transfer in a three-level superconducting circuit}

\author
{Antti Veps\"al\"ainen,$^{1}$ Sergey Danilin,$^{1}$ Gheorghe Sorin Paraoanu$^{1\ast}$\\
\\
\normalsize{$^{1}$ Low Temperature Laboratory, Department of Applied Physics, Aalto University School of Science}\\
\normalsize{P.O. Box 15100, FI-00076 AALTO, Finland}\\
\normalsize{$^\ast$To whom correspondence should be addressed; E-mail:  sorin.paraoanu@aalto.fi.}
}

\date{}

\newcommand{\bra}[1]{\langle #1|}
\newcommand{\ket}[1]{|#1\rangle}
\newcommand{\braket}[2]{\langle #1|#2\rangle}

\newcommand{\ex}[1]{\mathrm{e}^{#1}} % Exponential
\begin{document} 


\baselineskip24pt


\maketitle


\begin{sciabstract}
Adiabatic manipulation of the quantum state is an essential tool in modern quantum information processing. Here we demonstrate the speed-up of the adiabatic population transfer in a three-level superconducting transmon circuit by suppressing the spurious non-adiabatic excitations with an additional two-photon microwave pulse. We apply this superadiabatic method to the stimulated Raman adiabatic passage, realizing fast and robust population transfer from the ground state to the second excited state of the quantum circuit.
\end{sciabstract}

% In setting up this template for *Science* papers, we've used both
% the \section* command and the \paragraph* command for topical
% divisions.  Which you use will of course depend on the type of paper
% you're writing.  Review Articles tend to have displayed headings, for
% which \section* is more appropriate; Research Articles, when they have
% formal topical divisions at all, tend to signal them with bold text
% that runs into the paragraph, for which \paragraph* is the right
% choice.  Either way, use the asterisk (*) modifier, as shown, to

\section{Introduction}

The ability to accurately manipulate the state of quantum systems is one of the prerequisites for high-fidelity quantum information processing \cite{divincenzo2000physical}. The adiabatic control of quantum states is based on slowly modifiying the energy eigenstates of gapped systems; if the condition for adiabatic following is satisfied, the system remains in its instantaneous eigenstate at any moment in time. %The famous example of the process is the avoided crossing of the energy levels of the perturbed two-level system, resulting in excitation or relaxation of the system \cite{zener1932non}.
%Quantum control - the manipulation of a system such that it reaches a target state or that it follows a given path in the Hilbert space - is an essential tool of modern quantum information processing.
Techniques that are generically referred to as shortcuts to adiabaticity %\cite{Demirplak03,Demirplak05,Demirplak08,Berry09, Torrontegui13,PhysRevA.82.053403,Chen10}, also referred to as superadiabatic methods \cite{Giannelli14}, 
\cite{Torrontegui13}
aim at achieving faster operation times through a guided evolution of the system towards the desired final state, bypassing the restriction of the adiabatic theorem.

%typical for Rabi pulse manipulations by suppressing non-adiabatic excitations, while at the same time inheriting the robustness to errors of adiabatic protocols. 

For adiabatic quantum computing \cite{Farhi01}, quantum-annealing \cite{annealing,dwave}, and holonomic quantum computing \cite{fastholonomic,holonomic,nonabelianwallraff}, shortcuts to adiabaticity would be one important route to quantum advantage \cite{troyer}. These processes could provide a route toward quantum engines with increased efficiency \cite{del2014} and toward a better understanding of thermodynamical limits
\cite{PhysRevA.82.053403,B816102J,0295-5075-85-3-30008}. Furthermore, in multilevel quantum information processing \cite{Lanyon2009} shortcuts to adiabaticity can be employed for robust gates \cite{optimal_sa_stirap} and efficient initial state preparation.

Superadiabatic protocols (also called transitionless driving) \cite{Demirplak03,Demirplak05,Demirplak08,Berry09} are a type of shortcuts to adiabaticity based on counterdiabatic driving -- designed such that they suppress non-adiabatic excitations; in consequence, the system follows the instantaneous Hamiltonian eigenstate at any time during evolution. These protocols are universal, %independent on the state, 
and the robustness against errors is inherited from the corresponding adiabatic process. 
However, a major difficulty in implementing them stems from the fact that the superadiabatic control drive employs complex couplings with externally-controlled and stable Peierls phases \cite{Peierls:1933}. This is a key reason why the experimental implementation been restricted so far only to simple configurations, involving either two levels \cite{Bason12,Zhang13} or two control fields \cite{An2016,Du2016,Zhou2017}.

Here we demonstrate a superadiabatic protocol in the microwave regime using circuit quantum electrodynamics as the experimental platform \cite{cqed_strong_coupling_wallraff}. This is an important task in quantum control, where efficient and fast state preparation is needed as an initial step for many algorithms \cite{microwave_reset_wallraff,creating_dark_state_yang}. %To make use of the full potential of the superadiabatic concept, we demonstrate its validity for multilevel systems manipulated by several control fields through both amplitude and phase.
Our experiment employs a superconducting transmon circuit \cite{transmon_PRA2007}, where the first three states are addressed by microwave fields with the goal of populating the second excited state, while starting from the ground state and without populating the first excited state. We use three microwave pulses, of which two realize the stimulated Raman adiabatic passage (STIRAP) \cite{stirapfirst,stirap_review_vitanov_2017,stirap_ours}. The third one 
produces the counterdiabatic drive, which forces the system to follow its instantaneous eigenstate even though the adiabatic condition is  violated. This type of driving, called loop configuration \cite{Unanyan97}, results in an externally-controlled gauge-invariant phase and implements the superadiabatic STIRAP (saSTIRAP) protocol \cite{Chen10,Giannelli14}.

\begin{figure}[tbp]
\centering
\includegraphics[width = 1.0\textwidth]{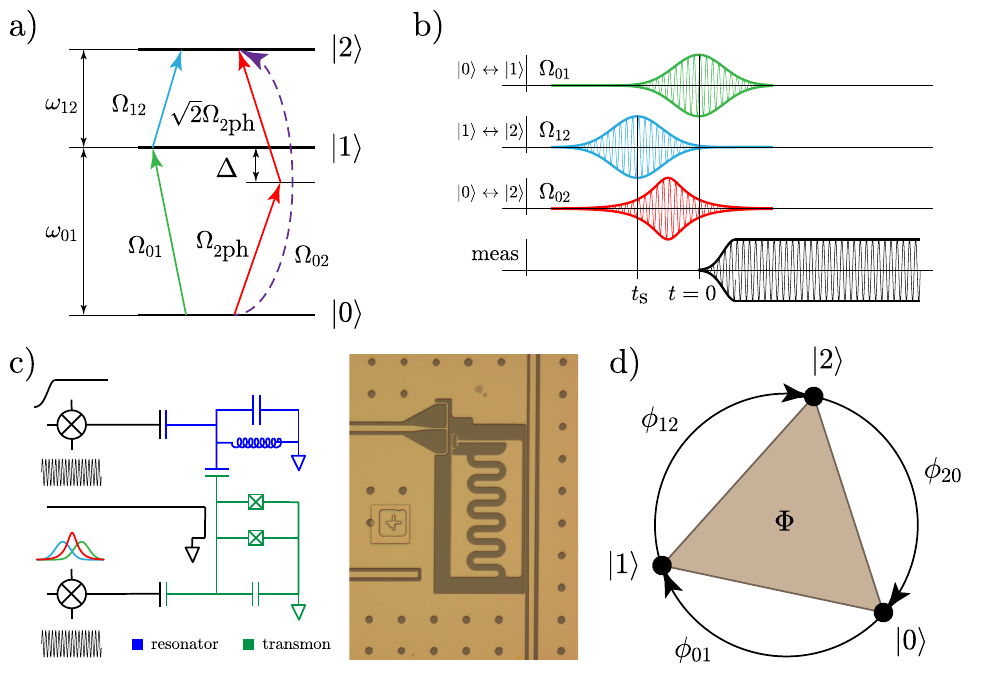}
\caption{{\bf Schematic of the experiment.}
a) Loop driving for saSTIRAP: a counterdiabatic drive with effective Rabi frequency $\Omega_{02}$ (dashed blue arrow) acts alongside the STIRAP sequence consisting of pulses $\Omega_{01}$ and $\Omega_{12}$, which are on-resonance with the transitions 0 - 1 and 1 - 2 respectively.
The two-photon process realizing the counterdiabatic correction is implemented by an off-resonant pulse (with detuning by $\Delta$ with respect to the 0 - 1  transition) which couples with strengths $\Omega_{\rm 2ph}$ and $\sqrt{2}\Omega_{\rm 2ph}$ into the corresponding transitions. b) saSTIRAP pulses (green, blue, red colours)in time-domain (schematic), as used in the experiment. After the protocol, a longer measurement pulse (black colour) is applied to the resonator. c) Schematic (including the IQ mixers used for driving and measurement) and optical image of the transmon. d) %In the synthetic space, the Hamiltonian describes a three-site plaquette with Peierls hopping and with total magnetic flux penetrating the plaquette 
Three-site plaquette representation with complex Peierls couplings, resulting in a gauge-invariant phase $\Phi = \phi_{01} + \phi_{12} + \phi_{20}$. 
}
\label{fig:schematic_saSTIRAP}
\end{figure}

%In the STIRAP protocol, the transitions of the three-level system are driven by microwave pulses \cite{stirap_ours}. 

For a three-level system in the ladder configuration, the resonant STIRAP Hamiltonian can be written as
\begin{equation}
  \label{eq:stirap_hamiltonian}
  H_0(t) = \frac{\hbar}{2}\left[\Omega_{01}(t)
    \ex{i\phi_{01}}
    \ket{0}\bra{1}
    + \Omega_{12}(t)\ex{i\phi_{12}}\ket{1}\bra{2} + {\rm h.c.} \right],
\end{equation}
where $\Omega_{01}(t)$ and $\Omega_{12}(t)$ describe the Rabi coupling of the microwave drive pulses to the transmon in the frame rotating with the drive frequencies. The drives have a Gaussian shape \cite{stirap_ours}
\begin{equation}
  \label{eq:stirap_amplitudes}
  \begin{aligned}
%\begin{aligned}
%\Omega_{01}(t) = \Omega\exp\left[-\frac{(t - t_{s}/2)^2}{2\sigma^2}\right], \\
%\Omega_{12}(t) = \Omega\exp\left[-\frac{(t + t_{s}/2)^2}{2\sigma^2}i\right], \\
  \Omega_{01}(t) &= \Omega_{01}\exp\left[-t^2 /(2\sigma^2)\right],\\
  \Omega_{12}(t) &= \Omega_{12}\exp\left[-(t - t_{\rm s})^2/(2\sigma^2)\right],
  \end{aligned}
\end{equation}
where $t_{\rm s}$ is the lag between the two pulses.
In the experiment we 
%A schematic of the experiment is shown in \ref{fig:schematic_saSTIRAP}a), b), and c).
%To create the matrix elements for the STIRAP sequence in Eq. Eq. \eqref{eq:hamiltonian_zero} we
employ two intermediate frequency microwave tones with externally-controlled phases $\phi_{01}$ and $\phi_{12}$, which are digitally mixed with the pulse envelopes $\Omega_{01}(t)$ and $\Omega_{12}(t)$ using an arbitrary waveform generator (see Supplementary Materials for details). The pulses are further mixed in an analog IQ-mixer with a local oscillator tone $\omega_{\rm LO}/(2\pi) = 6.92$ GHz to produce two signals that resonantly drive the 0\---1 and 1\---2 transitions of the three-level system at frequencies $\omega_{01}/(2\pi)=6.99$ GHz and $\omega_{12}/(2\pi) = 6.62$ GHz, see Fig. \ref{fig:schematic_saSTIRAP}. 
In the STIRAP protocol, the system follows adiabatically the dark state, defined as $|\rm D(t)\rangle =  \cos{\Theta (t)}e^{i \phi_{12}}\ket{0} - \sin{\Theta (t)}e^{-i \phi_{01}}\ket{2}$, where $\Theta (t) = \tan^{-1} [\Omega_{01}(t)/\Omega_{12}(t)]$ changes slowly from $0$ to $\pi/2$. This implies that the pulse driving the 1\---2 transition is counter-intuitively applied before the 0\---1 pulse, enabling the population to be transferred directly to the second excited state without exciting the intermediate state $|1\rangle$ at any time in-between. However, if the change in the amplitudes of the control signals is too abrupt, the adiabaticity is no longer fulfilled, resulting in a reduced transferred population and therefore limiting the fidelity of the process.

The spurious excitations of STIRAP can be cancelled using the superadiabatic method \cite{Berry09,Demirplak03,Demirplak05,Demirplak08}. The idea is to design a new control Hamiltonian which evolves the system through the adiabatic states given by the STIRAP Hamiltonian in Eq. \eqref{eq:stirap_hamiltonian}, even when the adiabatic condition is not fully satisfied \cite{Demirplak03}. The form of the counterdiabatic Hamiltonian can be found by reverse Hamiltonian engineering \cite{Berry09,Giannelli14,Chen10} %\cite{Chen10,Giannelli14,Fleischhauer99,Unanyan97}
(see Supplementary Materials for the derivation), requiring the addition of a third control pulse given by
\begin{equation}
H_{\rm cd}(t) = \frac{\hbar}{2}\left[\Omega_{02}(t)\ex{-i\phi_{20}}\ket{0}\bra{2} + {\rm h.c.}\right],\label{eq:cd}
\end{equation} 
with Rabi coupling
\begin{equation}
\Omega_{02}(t) = 2\dot{\Theta}(t),
\label{eq:omega02}
\end{equation}
 and a phase $\phi_{20}$ which must satisfy the relation $\phi_{01}+\phi_{12}+ \phi_{20} = - \pi/2$ \cite{spatial}. For the STIRAP pulse amplitudes given in Eq. \eqref{eq:stirap_amplitudes} and assuming $\Omega_{01} = \Omega_{12}$ the shape of the counterdiabatic pulse can be evaluated as \cite{Giannelli14}
\begin{equation}
\Omega_{02}(t) = -\frac{t_{\rm s}}{\sigma^2}\frac{1}{\cosh\left[-\frac{t_{\rm s}}{\sigma^2}(t - t_{\rm s}/2)\right]}.
\label{eq:omega02_2}
\end{equation}
% (here the notation $\phi_{20} = -\phi_{02}$ is used for the cyclic symmetry). 
 
To create the microwave pulse implementing the counterdiabatic Hamiltonian
%Eq. \eqref{eq:cd}
%$\langle 0 | H_{\rm cd} (t)| 2 \rangle = \langle 2 | H_{\rm cd} (t)| 0 \rangle^{*} = (\hbar /2) \Omega_{02}(t) \exp (-i \phi_{20})$ %with the amplitude and phase dependent on the other two tones used.
we use an additional microwave tone with frequency $\omega_{\rm 2ph} = (\omega_{01} + \omega_{12})/2$ and phase $\phi_{\rm 2ph}$, which couples into the 0\---1 and 1\---2 transitions with respective Rabi couplings $\Omega_{\rm 2ph}$ and  $\sqrt{2}\Omega_{\rm 2ph}$.  The frequency $\omega_{\rm 2ph}$ is chosen such that the two-photon resonance condition is satified. The factor $\sqrt{2}$ is a consequence of the almost harmonic energy level structure of the transmon circuit, which results in a higher dipole coupling for higher transitions \cite{transmon_PRA2007}. The low anharmonicity also leads to selection rules that prevent us from using a direct 0\---2 drive to implement the counterdiabatic Hamiltonian. The chosen drive frequency results in detunings $\pm \Delta$ from both the 0\---1 and 1\---2 transitions, $\Delta = \omega_{01} - \omega_{\rm 2ph} = (\omega_{01} - \omega_{12})/2$.
%The two-photon pulse couples to the 0\---1 and 1\---2 transitions with phase $\phi_{\rm 2ph}$ and Rabi couplings $\Omega_{\rm 2ph} \equiv \Omega_{\rm 2ph}^{01} \approx \frac{1}{\sqrt{2}}\Omega_{\rm 2ph}^{12}$, where the factor $\sqrt{2}$ comes from the harmonic approximation of the transmon.
%The two-photon drive is detuned from both the $0-1$ and $1-2$ transitions by $\pm \Delta$, which is sufficiently large to avoid parasitic excitations to state $|1\rangle$.
The two-photon driving generates an effective Rabi coupling $\Omega_{02}(t) = \sqrt{2}\Omega_{\rm 2ph}^2/(2 \Delta )$ and phase $\phi_{20} = - 2\phi_{\rm 2ph} - \pi$, which can be obtained from perturbation theory \cite{optimal_sa_stirap,James2007}. In addition, two-photon driving creates small ac-Stark shifts to all the energy levels, which appear as dynamic detunings of the drive frequencies from the transitions. We compensate for this effect by slightly tuning the phases of all the drive pulses during the evolution (see Methods for details).

\vspace{2mm}
\section{Results}

\begin{figure}[tb]
\centering
\includegraphics[width=1.0\textwidth]{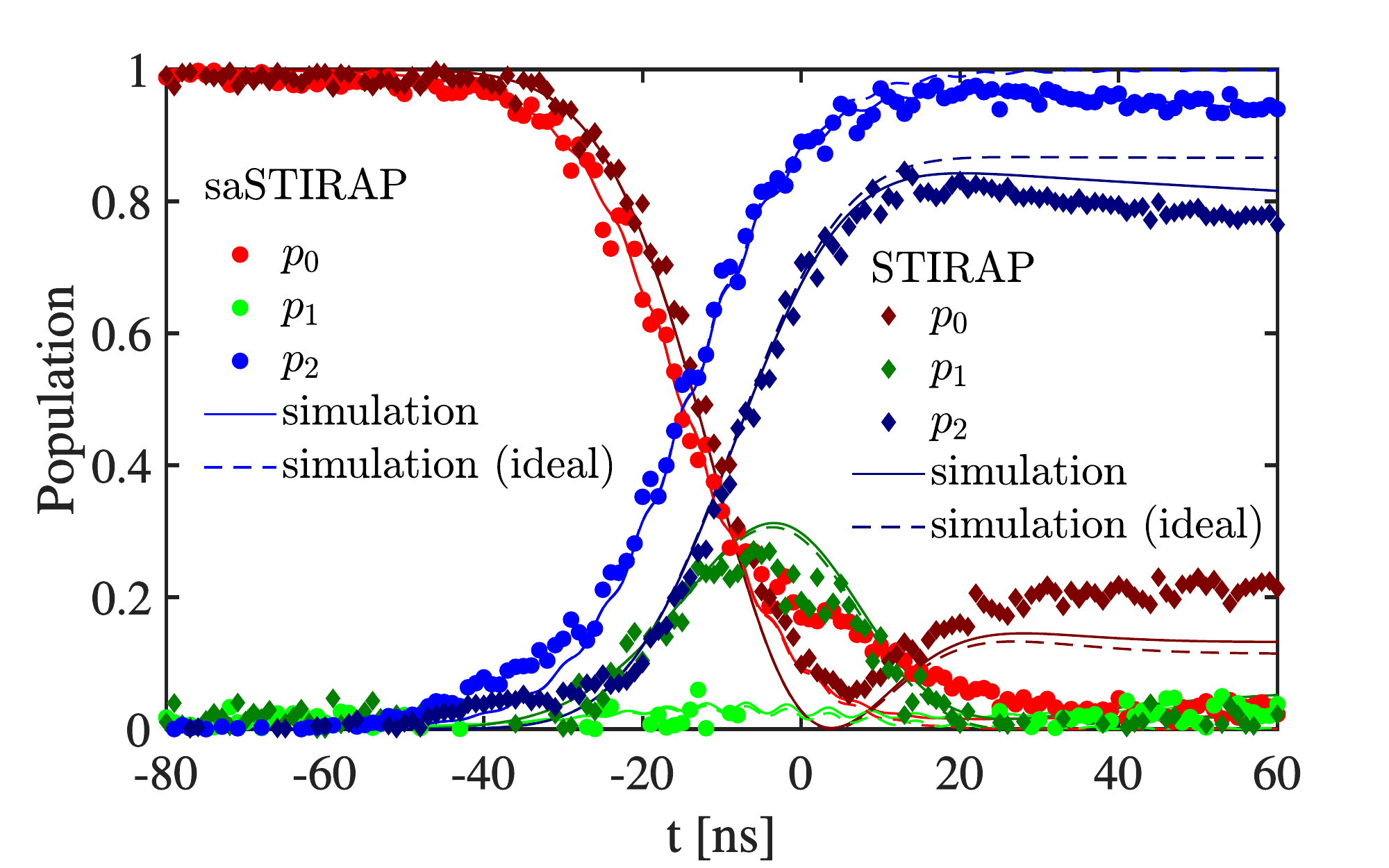}
\caption{{\bf Comparison between STIRAP and saSTIRAP.} Time evolution of the populations $p_0$, $p_1$, and $p_2$ during STIRAP (diamonds) and saSTIRAP (circles). The solid lines show the correspoding simulation, which includes decoherence. A simulation for the ideal case without decoherence is presented with dashed lines. The experiment was performed with the parameters $\Omega_{01} = \Omega_{12} =  25.5$ MHz, $t_{\rm s}/\sigma = -1.5$, and $\sigma = 20$ ns.}
\label{fig:time_sweep}
\end{figure}

\begin{figure}[tbp]
\centering
\includegraphics[width = 1.0\columnwidth]{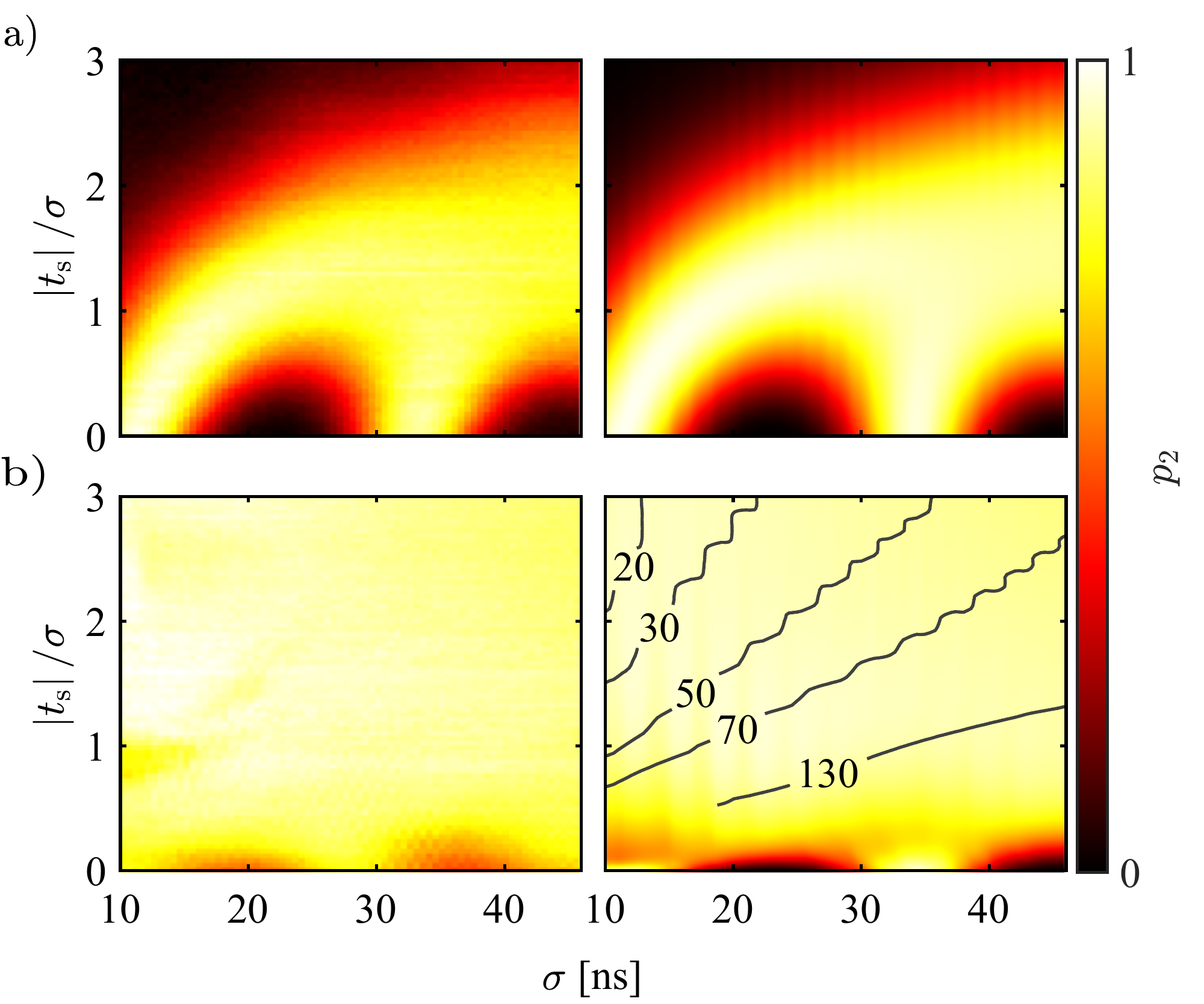}
\caption{{\bf Correction of the nonadiabatic losses with the saSTIRAP protocol.}
	a) The STIRAP process for various pulse widths $\sigma$ and normalized pulse separation $|t_{\rm s}|/\sigma$. We show here the population $p_2$ in the state $\ket{2}$ at amplitudes $\Omega_{01}/(2\pi) = \Omega_{12}/(2\pi) = 25.5$ MHz.
b) The saSTIRAP process for thes same parameters. The plots on the left are experimental data, while those on the right represent the results of simulation. Population $p_2$ for the corresponding saSTIRAP process. %with optimal $\tilde{\varphi}$. 
 The time $t_{\rm tr}^{0.9}$ (from 20 ns to 130 ns)  needed to achieve a population 
population $p_2 =0.9$
is shown with black lines.}
\label{fig:saSTIRAP_parameter_mapping}
\end{figure}

\subsection{Efficient transfer of population} To demonstrate that the superadiabatic protocol corrects for the non-adiabatic losses even when the adiabaticity condition for STIRAP is not satisfied, we experimentally compare the two methods in Fig. \ref{fig:time_sweep}. Here the peak STIRAP amplitudes $\Omega_{01}$ and $\Omega_{12}$ were chosen as $\Omega_{01}/(2\pi) = \Omega_{12}/(2\pi) = 25.5$ MHz, the separation of the two STIRAP pulses is $t_{\rm s}/\sigma = -1.5$, and the widths of the Gaussian pulse shapes are $\sigma = 20$ ns. During STIRAP there is a significant population in state $\ket{1}$ due to the violation of the adiabatic condition, which results in transitions between the instantaneous eigenstates of the system. Consequently, the population $p_2$ is only 0.8 after the pulses. In the saSTIRAP experiment there is almost no population in state $\ket{1}$ and $p_2$ reaches 0.96, which is very close to the ideal performance, demonstrating the power of the superadiabatic method. The result is supported by the numerical simulation, shown with solid lines (see Methods for details). The dashed lines show a simulation with the same parameters but without decoherence, resulting in $p_2 = 0.9997$ and confirming that most of the remaining losses in the saSTIRAP experiment are caused by the energy relaxation of the qutrit (with rates $\Gamma_{01} = 0.6$ MHz and $\Gamma_{12}= 0.83$ MHz, obtained by independent measurements).

In Fig. \ref{fig:saSTIRAP_parameter_mapping} we show the performance of the superadiabatic protocol for a wide range of STIRAP parameters.
%for a wide range of parameters. Here the peak STIRAP amplitudes $\Omega_{01}$ and $\Omega_{12}$ are kept constant at $\Omega_{01}/(2\pi) = \Omega_{12}/(2\pi) = 25.5$ MHz
We explore STIRAP and saSTIRAP in the parameter space $(t_{\rm s}, \sigma )$ by changing the width $\sigma$ and the normalized time-separation $|t_s|/\sigma$, as shown in Fig. \ref{fig:saSTIRAP_parameter_mapping}a). The optimal pulse separation for STIRAP is $t_{\rm s}/\sigma = -1.5$ \cite{stirap_review_vitanov_2017}. In the upper part of the plot the STIRAP fidelity is low because the separation of the pulses is too large, whereas for small $\sigma$ the adiabatic condition is not satisfied. STIRAP also fails for too small pulse separations; some high fidelity population transfer seen around $t_{\rm s} = 0$ in the experiment is not due to STIRAP but is driven by the holonomic gate studied in \cite{nonabelianwallraff,holonomic_gates_danilin}.
%in the left part of the figure the STIRAP pulse area is too small to satisfy the adiabatic condition, even though the pulse separation would be close to the optimum $|t_{\rm s}/\sigma| = 1.5$.
The experiment can be compared to a numerical simulation, which replicates the results accurately (right panel in the figure).
%In order to At each point the algorithm searches for and selects the optimal value of $\Phi = - 2\phi_{\rm 2ph} - \pi$.
%characterize the superadiabatic process, it is useful to introduce the parameter $|t_{\rm s}|/\sigma$ as the normalized pulse separation of the STIRAP pulses. 
%If the STIRAP amplitudes $\Omega_{01}$ and $\Omega_{12}$ are kept constant, we can obtain the entire experimentally accessible parameter space by varying $\sigma$ and the ratio $|t_s|/\sigma$, as shown in \ref{fig:saSTIRAP_parameter_mapping}.
%For this experiment we took $\Omega_{01}/(2\pi ) = 44$ MHz and $\Omega_{12}/(2\pi ) = 37$ MHz. %which are both much smaller than the qubit anharmonicity $2\Delta/(2\pi) = 282$ MHz.  
%For qubits with higher anharmonicity it would be advantageous to use even higher values for $\Omega_{01}$ and $\Omega_{12}$ in order 
%to improve the transfer efficiency. %Another option is to increase the width $\sigma$, but in practice this is limited by the decoherence rate. 
%From \ref{fig:saSTIRAP_parameter_mapping}a) we can see that typically STIRAP works well when the pulses are relatively close to each other, corresponding to $|t_{\rm s}|/\sigma = 1.5$.
The advantage of the saSTIRAP protocol over STIRAP is clearly demonstrated in Fig. \ref{fig:saSTIRAP_parameter_mapping}b), which shows high population transfer for all parameters used.

Next, we characterize the protocol by calculating the speed of transfer and comparing it to the quantum speed limit. The total time of the transfer is conventionally defined as the time lapse beween starting with 0.99 population in the ground state and achieving finally 0.9 in the second excited state \cite{Giannelli14}. This corresponds to initial and final mixing angles of $\Theta_{\rm i} = 0.03~\pi$ and $\Theta_{\rm f} = 0.4~\pi$, respectively. For calculating the quantum speed limit we use the Bhattacharyya bound \cite{Bhattacharyya83} for the subspace spanned by $|0\rangle$ and $|2\rangle$ under the maximal experimentally accessible two-photon Rabi drive $\Omega_{02}^{\rm max}/(2\pi) = 48$ MHz. We take the initial and the final states with the same populations as above, which results in
$T_{\rm QSL}^{0.9} =  2\arccos|\braket{\rm D(\theta_{\rm i})}{\rm D(\theta_{\rm f})}|/\Omega_{02}^{\rm max} \approx 7.7$ ns.
%This is the experimental value of $\Omega_{02}$ in saSTIRAP at $\sigma = 10$ ns and $t_{\rm s} = -30$ ns (the upper left corner in \ref{fig:saSTIRAP_parameter_mapping}), resulting in a quantum speed limit of $T_{\rm QSL}^{0.8} = 7$ ns.
The quantum speed limit can be compared to the transfer times for the saSTIRAP protocol which are shown by the overlayed solid lines in Fig. \ref{fig:saSTIRAP_parameter_mapping}b). The transfer times are the fastest ($T_{\rm saSTIRAP} \approx 2.0 T_{\rm QSL}$) in the upper left corner of the panels corresponding to $\sigma = 10$ ns and $|t_s/\sigma|$ = 3. However, as we approach that point, the STIRAP fidelity is also reduced and in consequence the population transfer occurs predominantly due to the counterdiabatic driving. Thus, the population transfer will start to be increasingly sensitive to the amplitudes of the pulses. In order to improve the robustness, the strength of the STIRAP part must be increased by reducing $t_{\rm s}/\sigma$, or by increasing $\sigma$, which leads to a reduction of transfer speed. The trade-off is important for the potential applications of the superadiabatic method and will be analyzed later in greater detail. 

% Next the relation between the adiabaticity and the r
%In order to understand the relation between the strength of the adiabatic part and the robustness which parameters to choose for optimal operation of the protocol we next analyze the relation of adiabaticity of the STIRAP part to the robustness of the population transfer.
% represent constant-value transfer times for the STIRAP and saSTIRAP protocols, and the dashed lines in the STIRAP simulation 
%show $p_2=0.8$. In STIRAP, this population level is reached only in the area delineated by the dashed line while in
%saSTIRAP the value $p_2$ is everywhere higher than 0.8.

%In that respect, it would seem most beneficial to apply the protocol with small $\sigma$ and large $t_{\rm s}$ in order to achieve fastest possible population transfer, but in that limit the behaviour of the protocol is no different from having the counterdiabatic pulse alone, without the STIRAP part. Next we demonstrate that in order to benefit from the robustness to the control parameters manifested by the adiabatic process, it is important that both diabatic and adiabatic contributions are present.

\begin{figure}[tbp]
\centering
\includegraphics[width = 1.0\columnwidth]{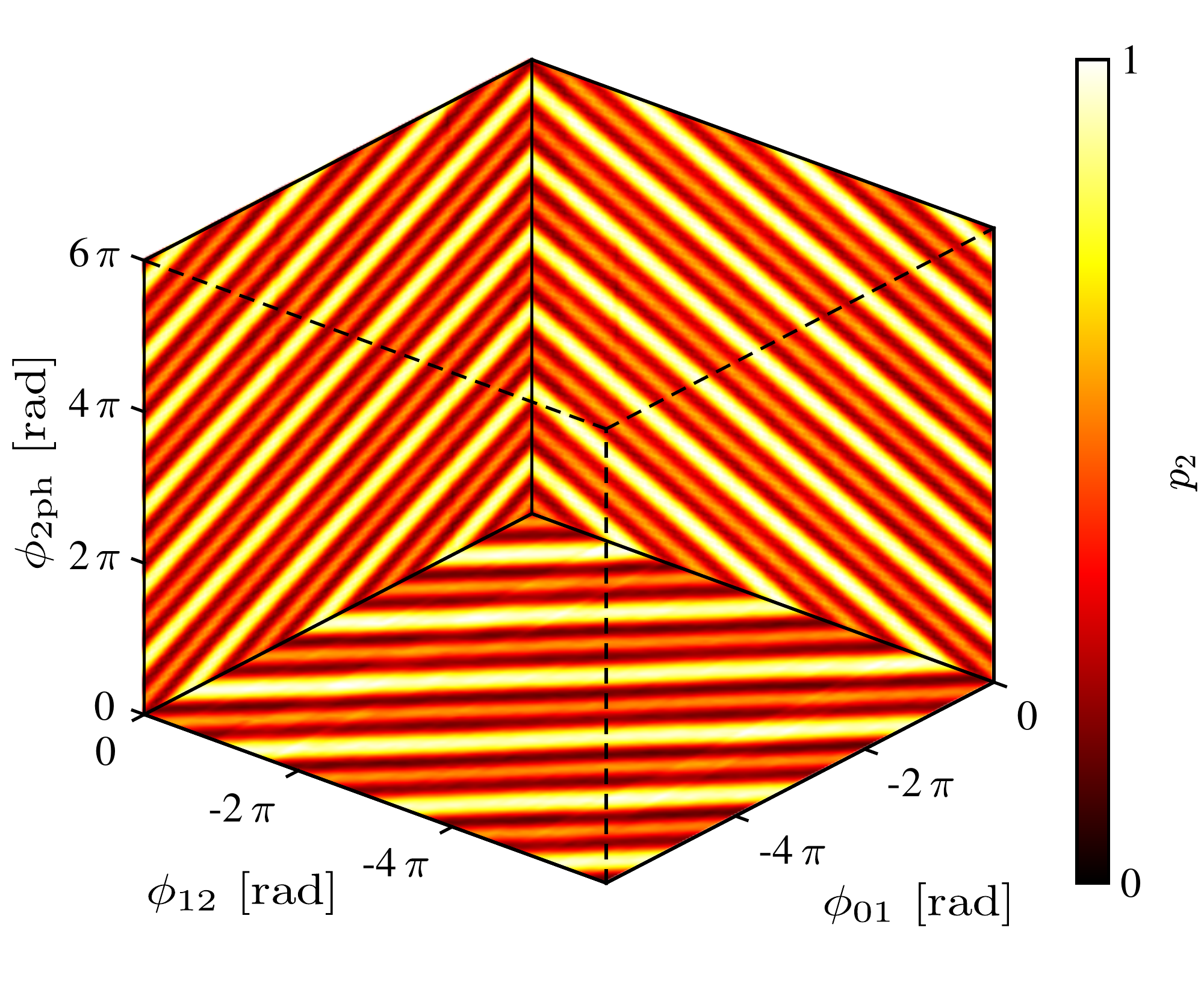}
\caption{{\bf Control of the system dynamics with the gauge-invariant phase.}
	The 3D plot shows lines of constant population $p_2$ in the orthogonal planes $(\phi_{12},\phi_{\rm 2ph})$ (with $\phi_{01}$ constant), $(\phi_{01},\phi_{\rm 2ph})$ (with $\phi_{12}$ constant) and $(\phi_{12},\phi_{01})$ (with $\phi_{\rm 2ph}$ constant). Similarly to lattice gauge theories, the phase $\Phi$ is a gauge-invariant quantity. In this geometrical representation, the relation $\phi_{01}+\phi_{12} - 2\phi_{\rm 2ph} - \pi = \Phi$ can be visualized as tilted planes that slice the axes.
Note also that the periodicity along the $\phi_{\rm 2ph}$ axis is twice that of the periodicity  along the axes $\phi_{01}$ and $\phi_{12}$. The experimental parameters used were $\Omega_{01}/(2\pi) = \Omega_{12}/(2\pi) = 25.5$ MHz, $t_{\rm s} = -30$ ns, and  $\sigma = 20$ ns. }
%and $\tilde{\Omega}_{01} = \sqrt{2}\tilde{\Omega}_{12} = 38$ MHz.}
\label{fig:saSTIRAP_full_measurement}
\end{figure}

\subsection{Gauge-invariant phase} Loop driving with complex couplings between each pair of states results in a nontrivial synthetic gauge structure on the triangular plaquette formed by the three states, previously studied theoretically in \cite{spatial,synthetic_arbitrary_gauge_qutrit_vitanov}; related schemes have been proposed for cold atom lattices in \cite{Lewenstein2014}. See Fig. \ref{fig:schematic_saSTIRAP}d) for a simple illustration.

%induced by the counterdiabatic term, as anticipated in \ref{fig:schematic_saSTIRAP}d). 
%We note that the Hamiltonian in Eq. (\ref{eq:full_hamiltonian}) describes three simultaneous rotations in the three subspaces $0-1$, $1-2$, and $0-2$ around the vectors $\hat{{\bf n}}_{kl}$. In each of the subspaces $(k,l)$, the action of the Hamiltonian is analogous to that of a spin-1/2 particle in a magnetic field of magnitude $\Omega_{kl}$ and direction $\hat{{\bf n}}_{kl}$. In one subspace $(k,l)$ it is possible to rotate arbitrarily the axis to align one of them along $\hat{{\bf n}}_{kl}$. It is possible to do this in two subspaces simultaneously, but crucially, one cannot rotate arbitrarily all the three vectors $\hat{{\bf n}}_{kl}$. Formally, by applying a unitary local gauge transformation of the form $U = e^{-i \chi_0} |0\rangle \langle 0| + e^{-i \chi_1} |1\rangle \langle 1|+ e^{-i \chi_2} |2\rangle \langle 2|,$ where $\chi_0$, $\chi_1$, and $\chi_2$ are arbitrary phases, 
%one obtains a Hamiltonian with a similar structure to Eq. (\ref{eq:full_hamiltonian}) with different angles $\phi_{kl}'$ (see Supplemetary Note 2); however, these new angles are not independent of each other, but they must satisfy the constraint $\phi_{01}' + \phi_{12}' + \phi_{20}' = \phi_{01} + \phi_{12} + \phi_{20} = \Phi$. Thus, by performing this transformation we can always eliminate two of the phases but the third one will be constrained by the value of the gauge-invariant quantity $\Phi$. 

In Fig. \ref{fig:saSTIRAP_full_measurement} we demonstrate experimentally that in a three-level transmon the gauge-invariant phase $\Phi = \phi_{01} + \phi_{12} + \phi_{20}$ fully determines the dynamics, once the pulse amplitudes are fixed. We show the population transferred onto the state $\ket{2}$, when one of the phases $\phi_{01}$, $\phi_{12}$, or $\phi_{\rm 2 ph}$ is kept constant and the other two change. The measurement is taken at $t=20$ ns after the 0 \--- 1 drive pulse reaches its maximum. The two-photon pulse is realized such that Eq. \eqref{eq:omega02} is satisfied.
%when the transferred is typically completed. most of the population has been  transferred to state $\ket{2}$. 
The experiment demonstrates that the population $p_2$ depends only on $\phi_{01} + \phi_{12} + \phi_{20} = \Phi$ and not on each phase separately \cite{spatial}. This allows us to choose the gauge $\phi_{01} = \phi_{12} = 0$ and use $\phi_{20} = \Phi = -2\phi_{\rm 2ph}-\pi$ as the externally controlled gauge-invariant phase.
%The $\pi$-periodicity of the population transferred as a function of the phase $\tilde{\varphi}$ of the two-photon drive pulse. In contradistinction, a sequential process (where we populate the first excited state, then transfer to the second excited state) should display a $2 \pi$ periodicity in the single-photon drive phase. This demonstrates the fully quantum-coherent nature of the process.The figure demonstrates that the choice of the three phases is not independent from each other. We can understand their relation by looking at Hamiltonian in Eq. Eq. \eqref{eq:full_hamiltonian}, which 
%A convenient choice of gauge is $\phi_{01}' = 0$, $\phi_{12}' = 0$, and $\phi_{20}'= - \phi_{02}' = \Phi$, which leads to the following structure for Eq. (\ref{eq:full_hamiltonian}),

In this gauge the full Hamiltonian of the system reads
\begin{equation}
  \begin{aligned}
    H(t) =& \frac{\hbar}{2}\left[\Omega_{01}(t)\ket{0}\bra{1} + \Omega_{12}(t)\ket{1}\bra{2} \right. \\
    & \left. + \Omega_{02}(t)\ex{-i\Phi}\ket{0}\bra{2}+{\rm h.c.} \right],
  \end{aligned}
\end{equation}
%% \begin{equation}
%% H(t) = \frac{\hbar}{2} \Omega_{01}(t) \sigma_{01}^{x} + \frac{\hbar}{2} \Omega_{12}(t) \sigma_{12}^{x}+ \frac{\hbar}{2} \Omega_{02}(t) \hat{{\bf n}}_{\Phi}\cdot \bm{\sigma}_{02}, \label{eq:full_hamiltonian_standard}
%% \end{equation}
%% where $\hat{{\bf n}}_{\Phi} = (\cos \Phi , \sin \Phi )$, see also the Supplemental Material \cite{supplemental_material}.
 thus simplifying the problem significantly (see also the Supplementary Materials).

\begin{figure}[tbp]
\centering
\includegraphics[width = 1.0\textwidth]{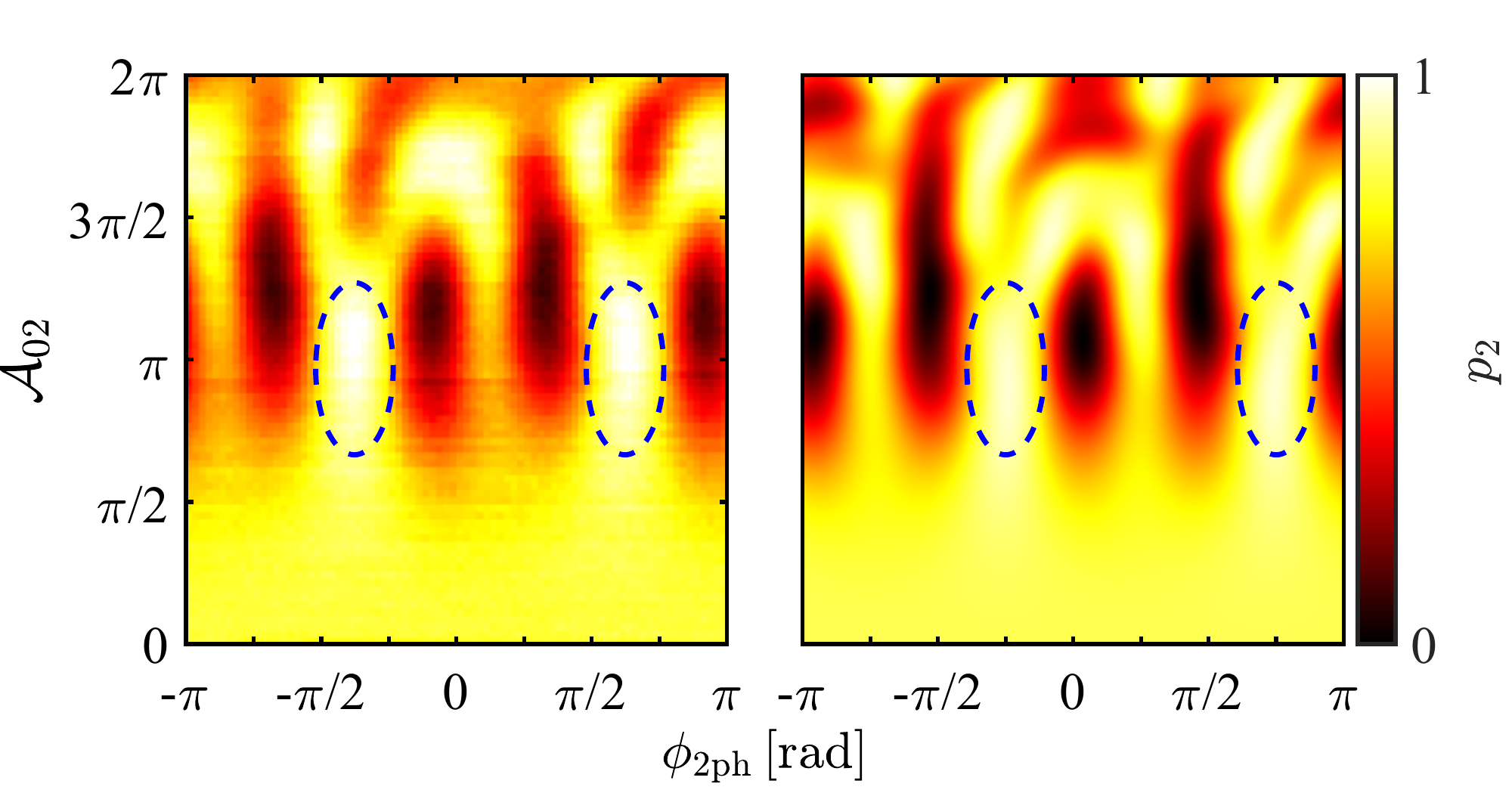}
\caption{{\bf Robustness of saSTIRAP against variations in the counterdiabatic pulse parameters.}
Dependence of $p_2$ on the area of the counterdiabatic pulse and on the gauge-invariant phase. 
The left panel shows the experimental data while the right panel shows the corresponding simulation.
The experimental parameters were $t_{\rm s} = -30$ ns, $\sigma = $ 20 ns, and $\mathcal{A} = 4.2\pi$. Pure STIRAP is achieved at $\mathcal{A}_{02} =0$. The areas where saSTIRAP is robust against changes in parameters $\mathcal{A}_{02}$ and $\phi_{\rm 2ph}$ is shown as blue dashed-line ellipses.
%b) Phase dependence of the superadiabatic process as a function of the STIRAP amplitude with the experiment in the left panel and the simulation in the right panel. $\mathcal{A} = 0$ corresponds to the case without STIRAP and is thus completely phase independent. The experiment was performed with $t_{\rm s} = -37.5$ ns, $\sigma = 25$ ns, and $\mathcal{A}_{02} = 0.81\pi$.
%Some parameters for a) $k=2.45$, $\sigma = 25$ ns, $\Omega_{01} = 44$ MHz, $\Omega_{12} = 36.8$ MHz, $A_{\rm eff} = 5.5\pi$.
% b) ($\tilde{\Omega}_{02} := \tilde{\Omega}_{01}$ = 50 MHz, $\Omega_{02} = 12.5$ MHz $\hat{=} \pi$), $k = 1.5$,$\sigma = 25$ ns}
}
\label{fig:areaandphase}
\end{figure}

\subsection{Robustness properties} A major advantage of STIRAP is its robustness against errors in the amplitudes of the drive fields. The crucial question is whether this robustness extends to the amplitude of the counterdiabatic field, as for the practical applications an error-resilient saSTIRAP would provide a similar improvement over the non-adiabatic methods. First we define the area of the counterdiabatic pulse 
\begin{equation}
\mathcal{A}_{02} = \int_{-\infty}^{\infty} {\mathrm d}t\, \Omega_{02}(t),
\label{eq:two_photon_area}
\end{equation}
and the STIRAP pulse area as
\begin{equation}
\mathcal{A} = \int_{-\infty}^{\infty} {\mathrm d}t\,\sqrt{\Omega_{01}^{2}(t) + \Omega_{12}^{2}(t)} \label{eq:stirap_area},
\end{equation}
which is the measure of adiabaticity of STIRAP according to the global adiabatic condition $\mathcal{A} \gg \pi/2$ \cite{stirap_review_vitanov_2017}. 
In Fig. \ref{fig:areaandphase} we present the population of state $\ket{2}$, $p_2$ at various values of the area and the phase of the counterdiabatic pulse.
We note that inside the regions outlined with blue dashed-line ellipses the pulse areas $\mathcal{A}_{02}$ are close to $\pi$, as expected from Eq. \eqref{eq:omega02_2}, and the population $p_2$ is high. For the parameters $(\mathcal{A}_{02},\phi_{\rm 2ph})$ inside these ellipses, $p_2$ is a rather slow-varying function of $\mathcal{A}_{02}$, indicating that saSTIRAP is indeed robust against errors in the area of the counterdiabatic pulse.
Outside these ellipses the transfer is not robust with respect to changes in  $\mathcal{A}_{02}$. In the right panel we present the  numerical simulation, which matches the pattern seen in the experiment quite well. From the simulation we can also see that the maximum transfer occurs around an optimal phase which is very close to the ideal $\phi_{\rm 2ph} = -\pi/4 + n\pi$. In the experiment, a small shift exists in the phases due to the phase imbalance of the IQ-mixer (see Fig. 1c)) used to combine the driving pulses (more details available in the Supplementary Materials).

\begin{figure}[tbp]
	\centering
	\includegraphics[width = 1.0\columnwidth]{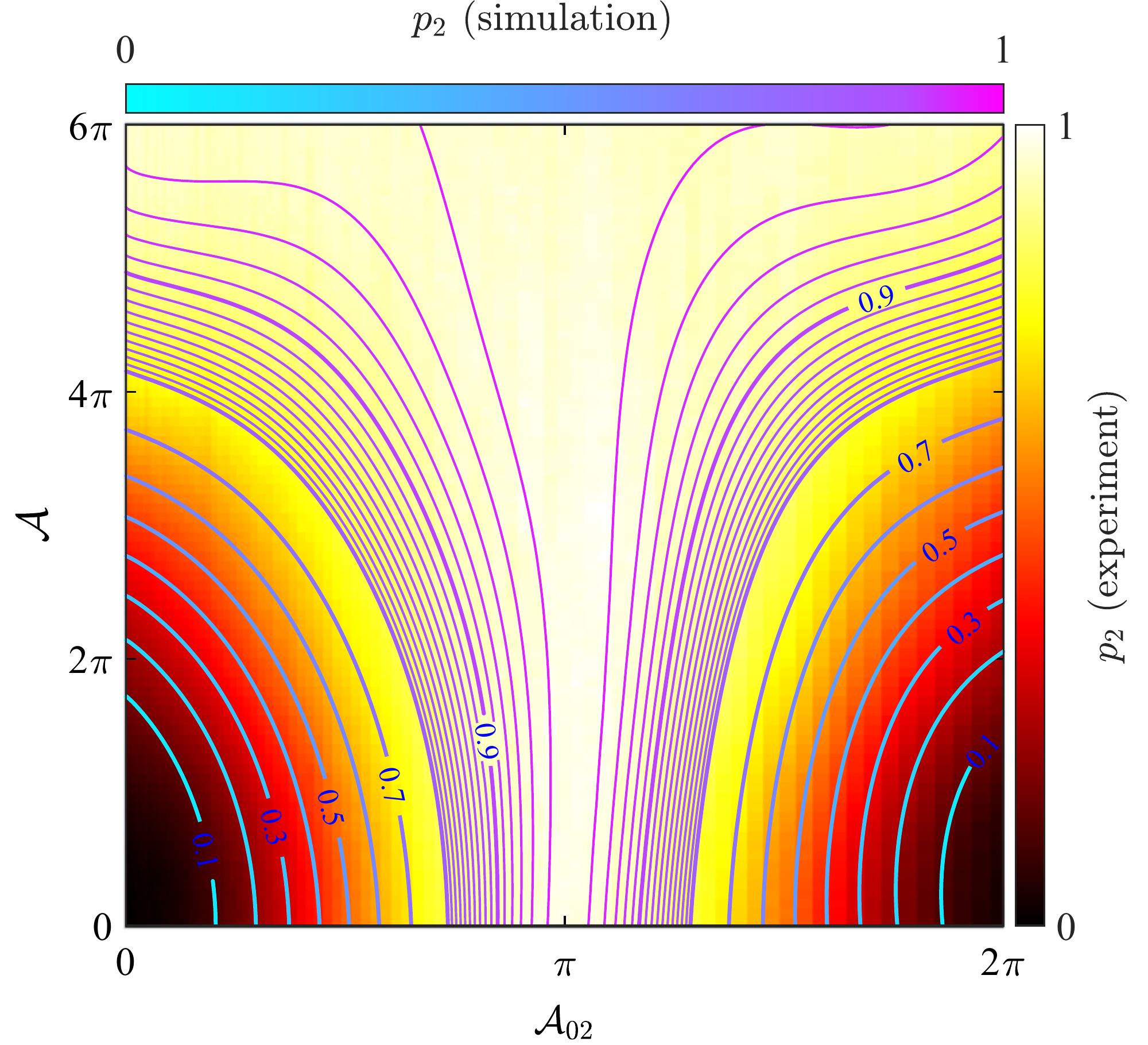}
    \caption{{\bf Comparison between saSTIRAP and nonadiabatic population transfer.}
	Dependence of the transferred population $p_2$ on the STIRAP pulse area $\mathcal{A}$
		as defined in Eq. \eqref{eq:stirap_area}, 
		and on the two-photon pulse area $\mathcal{A}_{02}$
		from Eq. \eqref{eq:two_photon_area}. To illustrate the agreement with theory, 
		we also show the iso-population lines (from 0.1 to 0.8 in steps of 0.1 and from 0.8 to 1.0 in steps of 0.01) obtained from the simulations. 
    %	showing agreement with the data and delineating the same region of high transfer as that obtained from experiment.
		%The solid black line corresponding to a population on state $|2\rangle$ of $P_{|2\rangle}= 0.4$ is shown in order to delineate the region where transfer occurs. The dotted lines show the values of the $\mathcal{A}_{02}$ area leading to the same population in the absence of the STIRAP pulse.
	In this experiment, the maximum Rabi couplings of the STIRAP process were $\Omega_{01}/(2\pi) = \Omega_{12}/(2\pi) = 40$ MHz. Similarly, the two-photon pulse amplitude was varied from zero to $\Omega_{\rm 2ph}/(2\pi) = 77$ MHz. The horizontal axis with $\mathcal{A} = 0$ corresponds to two-photon Rabi driving wheras the vertical axis with $\mathcal{A}_{02} = 0$ corresponds to standard STIRAP.
        In the experiment the STIRAP pulse separation was $t_{\rm s} = -30$ ns and the pulse width was $\sigma = 20$ ns.
		%b) Shows the interval $[\mathcal{A}_{02}]_{P_{|2\rangle} \geq 0.4}$  between the equal-population lines $P_{|2\rangle}= 0.4$ from a). Up to a critical value  $\mathcal{A} \approx 2\pi$ 
		%the transfer is driven by the two-photon pulse only; for $\mathcal{A} \geq 2\pi$, STIRAP starts to work and the width of the interval increases considerably, demonstrating robustness with respect to the two-photon area pulse. 
%		b) Transfer efficiency of saSTIRAP when the STIRAP pulse width $\sigma$ and the counterdiabatic pulse area are varied. The 		orange circles are values obtained in 5 different saSTIRAP experiments where 
%		the value $p_2 = 0.55$ was obtained, while the solid lines are simulations (with that for $p_2 =0.55$ thickened). The yellow circle is another saSTIRAP experiment, which yielded $p_2=0.77$.
%		The dashed line (simulation) corresponds to the case where $p_2 =0.55$ is obtained by applying only the counter-adiabatic pulse. 
		%The case where the counterdiabatic pulse area is zero (left vertical axis) corresponds to pure STIRAP. 
%		The amplitudes of the STIRAP pulses were $\Omega_{01}/(2\pi) = 25$ MHz and $\Omega_{12}/(2\pi ) = 16$ MHz.%, $A_{\rm eff} = 3.1\pi$, $\kappa = 1.5$, $\sigma = 25$ ns}
	}
	\label{fig:robustness}
\end{figure}

We can compare saSTIRAP with the direct non-adiabatic process by plotting (see Fig. \ref{fig:robustness}) the population of the second excited state as a function of $\mathcal{A}$ and $\mathcal{A}_{02}$. The phase $\phi_{\rm 2ph}$ is chosen such that the maximum population in state $|2\rangle$ is obtained for every value of the STIRAP area $\mathcal{A}$. When only the counteradiabatic pulse is present ($\mathcal{A}=0$) the transfer occurs only in a narrow range of $\mathcal{A}_{02}$ values around $\pi$. When the area of the STIRAP pulses is increased (at approximately $\mathcal{A} \approx 2\pi$), the range of values of $\mathcal{A}_{02}$ where the transfer occurs enlarges significantly. This demonstrates the advantage that the superadiabatic method offers: it has better fidelity than the adiabatic protocol and it is more robust against the variation of $\mathcal{A}_{02}$ than a raw $\pi$ pulse. Theoretically, the fidelity of STIRAP approaches unity only in the limit of infinite pulse area (the adiabatic condition is fully satisfied), whereas ideal saSTIRAP has unit fidelity for all the values of the STIRAP pulse area.

Fig. \ref{fig:robustness} also demonstrates that, even though the maximal effective two-photon coupling is smaller than the direct 0\---1 and 1\---2 couplings, it does not severely restrict the speed of the method, because the optimal two-photon pulse area $\mathcal{A}_{02} = \pi$ is usually significantly smaller than the STIRAP area $\mathcal{A}$ required to provide the demanded robustness. This is also an advantage over rapid adiabatic passage \cite{malinovsky2001general,Arkhipkin2003}, where a much stronger two-photon pulse on the 0\---2 transition would be needed.

\section{Methods}
  %\setcounter{subsection}{0}
  %\subsection{Circuit quantum electrodynamics}
  %{\color{red} Do you think that we need to tell about basics of circuitQED here?}
\subsection{Three-level quantum tomography}
The state of the qutrit is obtained by three-level quantum tomography, where the diagonal elements of the density matrix are calculated from the measured 
IQ trajectories \cite{three_level_tomography} of the cavity. After suitable averaging to reduce the noise, we obtain
\begin{equation}
r_{\rm meas}(\tau) = \sum_{i = 0,1,2} p_i r_i(\tau),
\end{equation}
written here as a linear combination of calibration traces $r_i(\tau)$ corresponding to states $\ket{0}$, $\ket{1}$ and $\ket{2}$ with weight factors $p_0$, $p_1$, and $p_2$, which give the occupation probability of each state. Here $\tau$ is the time from the beginning of the measurement pulse. Using the least squares fitting method, we can find the most likely occupation probabilities for the three-level system. Relaxation inevitably influences the calibration traces and must be compensated. We do this correction by modifying the calibration to include some contribution from the lower states, described by errors $\zeta_{ij}$ with $i < j$. The measured trajectory $r_j$ is then given by 

\begin{equation}
r_j(\tau) = \left(1 - \sum_{i<j} \zeta_{ij}\right)\tilde{r}_j(\tau) + \sum_{i<j} \zeta_{ij}\tilde{r}_i(\tau),
\end{equation}
with $\tilde{r_i}(\tau)$ describing the unknown ideal responses of state $\ket{i}$.
%makes them non-ideal, and effectively results in higher populations for states $\ket{1}$ and $\ket{2}$. 
%We compensate for that error by modifying the calibration trajectories by assuming that the each calibration trajectory has some contribution from the lower states, described by $\epsilon_{ij}$ with $i < j$. 
From the above equation the ideal responses can be solved iteratively, yielding
%% \tilde{r}_0 &= r_0, \\
%% \tilde{r}_1 &= \frac{r_1 - \epsilon_{01}*r_0}{1-\epsilon_{01}}, \\
%% \tilde{r}_2 &= \frac{r_2 - \epsilon_{02}r_0 - \epsilon_{12}\frac{r_1 - \epsilon_{01}r_0}{1-\epsilon_{01}}}{1-\epsilon_{02}-\epsilon_{12}}.
\begin{equation}
\begin{aligned}
\tilde{r}_0 (\tau ) &= r_0 (\tau ), \\
\tilde{r}_1 (\tau )&= \frac{r_1 (\tau )- \zeta_{01}\tilde{r}_0 (\tau )}{1-\zeta_{01}}, \\
\tilde{r}_2 (\tau )&= \frac{r_2 (\tau )- \zeta_{02}\tilde{r}_0 (\tau )- \zeta_{12}\tilde{r}_1 (\tau )}{1-\zeta_{02}-\zeta_{12}}.
\end{aligned}
\end{equation}
We use $\zeta_{01} = 0.01$, $\zeta_{12} = 0.01$, and $\zeta_{02} = 0.02$, which are obtained by comparing a reference Rabi experiment to a corresponding simulation with known energy relaxation rates.

\subsection{Dynamical phase correction}
The off-resonant two-photon driving produces parasitic ac-Stark shifts of the energy levels, which we compensate for by using dynamically adjusted phases. Following \cite{optimal_sa_stirap}, the ac-Stark shifts can be calculated from the second order perturbation theory as $\tilde{E}_n(t) = E_n + \bra{n}V(t)\ket{n} + \sum_{k\neq n} \frac{\bra{k}V(t)\ket{n}}{E_n - E_k}$, where $V(t)$ consists of the off-diagonal elements of the two-photon drive Hamiltonian $V = \hbar\Omega_{\rm 2ph}(t)\left(\ket{0}\bra{1}\ex{i\phi_{\rm 2ph}} + \sqrt{2}\ket{1}\bra{2}\ex{i\phi_{\rm 2ph}} + {\rm h.c}\right)/2$ in the frame rotating with the drive. The energies $E_n$ are the detunings of the drive from the 0\---1 and 1\---2 transitions, $E_0 = 0$, $E_1 = \hbar\Delta$, and $E_2 = 0$. %and $\lambda$ describes the increased coupling of the drive to the higher transitions. 
The resulting ac-Stark shifts $\epsilon_{n,k} = \tilde{E}_k - \tilde{E}_n - (E_k - E_n)$ are $\epsilon_{01}(t) = \hbar|\Omega_{\rm 2ph}|^2 /\Delta$, $\epsilon_{12} = -5\hbar |\Omega_{\rm 2ph}|^2/(4\Delta)$ and $\epsilon_{0,2} = -\hbar|\Omega_{\rm 2ph}|^2/(4\Delta)$. In order to compensate for the shifts in the energy levels, we dynamically modify the phases of all the three drives as $\phi_{nk}(t) \rightarrow \phi_{nk} + \int_{-\infty}^t \mathrm{d}t\, \epsilon_{nk}(t)/\hbar$. As a result, the frequencies of the drives match the ac-Stark shifted qutrit transition frequencies at all instants of time.

\subsection{Numerical simulations}
The system is modelled with the Hamiltonian
\begin{equation}
  H_{\rm sim}(t) = H_0 + \hbar\Omega_{\rm 2ph}(t)/2\left(\ket{0}\bra{1}\ex{i(\phi_{\rm 2ph}(t) - \Delta t)} + \sqrt{2}\ket{1}\bra{2}\ex{(i\phi_{\rm 2ph}(t) + \Delta t)} + {\rm h.c}\right)
\end{equation}
in the frame rotating with the STIRAP drives. Here $H_0$ is the STIRAP Hamiltonian given in Eq. \eqref{eq:stirap_hamiltonian} and the evolution of the system is solved from the Lindblad master equation $\dot{\rho}(t) = -i[H_{\rm sim}(t), \rho(t)]/\hbar +\sum_{i={0,1}}\Gamma_{i,i+1}\left(\ket{i}\bra{i+1}\rho(t)\ket{i+1}\bra{i} - \frac{1}{2}\left(\ket{i}\bra{i}\rho(t) + \rho(t)\ket{i}\bra{i}\right)\right)$,
where $\rho(t)$ is the density matrix of the system and $\Gamma_{i,i+1}$ are the energy relaxation rates (obtained by independent qubit characterization measurements).
%The pure dephasing rates were measured to be $\Gamma_\phi^{01} \approx \Gamma_\phi^{12} \approx 0.1$ MHz $\ll \Gamma_{i,i+1}$ and are thus neglected in the simulation.

%\showmatmethods{}

\section{Conclusions}
We have demonstrated a speed up of population transfer in STIRAP by introducing an additional counterdiabatic two-photon control pulse that corrects for non-adiabaticity. We have shown that the population transfer is determined by the pulse amplitudes and by a gauge-invariant phase. %The counterdiabatic pulse acts on the $\et{0} - \ket{2}$ transition, which is a forbidden direct transition in a transmon. We circumvent the problem by using a two-photon process, which effectively drives the desired transition. 
We have characterized the robustness properties of this superadiabatic process with respect to the counterdiabatic Hamiltonian, and we have evaluated the corresponding trade-off
with respect to the speed of the process.

\bibliography{ref_saSTIRAP}

\bibliographystyle{Science}

\section*{Acknowledgments}
This work used the cryogenic facilities of the Low Temperature Laboratory at Aalto University. We acknowledge financial support from FQXi, V\"aisal\"a Foundation, the Academy of Finland (project 263457), the Center of Excellence ``Low Temperature Quantum Phenomena and Devices'' (project 250280) and the ``Finnish Center of Excellence in Quantum Technology'' (project 312296).

AV and SD performed the experiments; SD fabricated the sample, and AV analyzed the results. AV wrote the manuscript together with GSP, with additional contributions from SD. GSP supervised the project. All the authors contributed to the planning of the experiments.

The authors declare that they have no competing interests. All data needed to evaluate the conclusions in the paper are present in the paper and/or the Supplementary Materials. Additional data available from authors upon request.

%Here you should list the contents of your Supplementary Materials -- below is an example. 
%You should include a list of Supplementary figures, Tables, and any references that appear only in the SM. 
%Note that the reference numbering continues from the main text to the SM.
% In the example below, Refs. 4-10 were cited only in the SM.     

\newpage

\section*{Supplementary Materials}

\subsection*{Experimental setup and sample}

%The superconducting QED circuit \cite{transmon_PRA2007} consists of a capacitively-shunted Cooper pair box (a transmon), with tunable energy level separation under the application of an external magnetic field. The qubit can be excited by applying a continuous wave or pulsed microwaves. The readout system is realized as a $\lambda /4$ coplanar waveguide cavity (resonator) which allows the quantum non-demolition measurement of the state of the transmon \cite{cqed_strong_coupling_wallraff,dispersive_readout_bianchetti}. A picture of the sample is shown in \ref{fig:sample_design}a),b), and a schematic of the measurement setup at room temperature and in the dilution refrigerator is presented in \ref{fig:sample_design}c).

\begin{figure}[!ht]
	\centering
	\includegraphics[width=0.7\columnwidth]{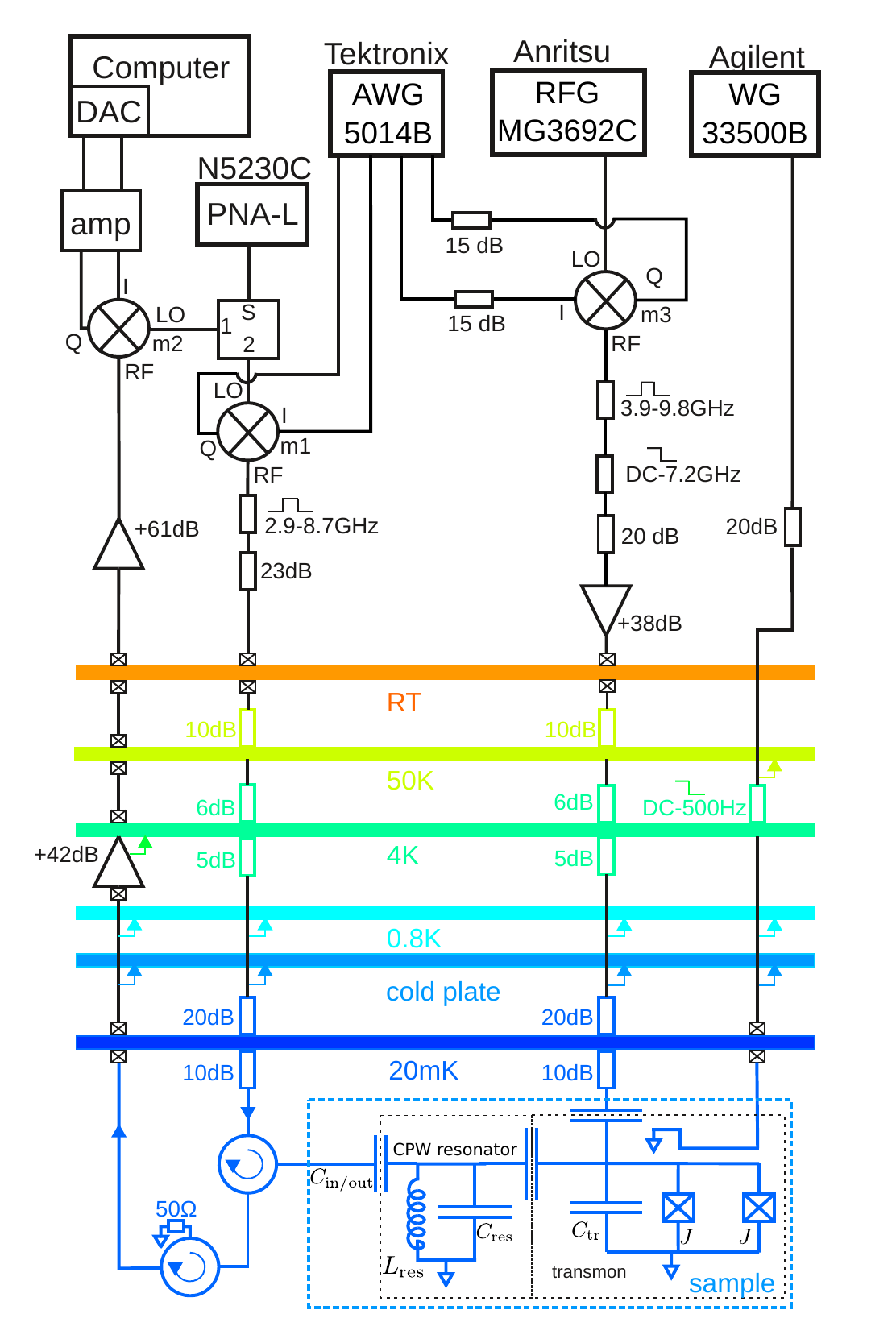}
	\caption{{\bf Electronics, cryogenics, and sample schematic.}
		The sample consists of a transmon device coupled to $\lambda/4$ coplanar waveguide resonator. It is placed at the mixing chamber of a dilution refrigerator and controlled by room-temperature electronics.}
	\label{fig:saSTIRAP_electronics}
\end{figure}

A transmon is an artificial atom with energy levels $\hbar\omega_{j}$, yielding transition frequencies
$\omega_{j,j+1} = \omega_{j+1} - \omega_{j}$, and can be modelled as an anharmonic oscillator. With standard notations in the field of superconducting devices, we denote by $E_{J}$ the Josephson energy of the transmon and by $E_{\rm C}$ the total charge energy (including the shunt capacitor). The anharmonicity is $\hbar \omega_{12}-\hbar \omega_{01} \approx - E_{\rm C}$ in the asymptotic limit $E_{\rm J} \gg E_{\rm C}$.
To manipulate the state of the transmon we apply  microwave drive signals
either to the gate or to the coupled transmission line cavity. If the frequency of the drive %$\omega_{j,j+1}^{(\Omega )}$
matches the transition frequency between two states of the system, the system goes through Rabi oscillations, resulting in transfer of population between the two states.

The experimental setup including the schematic of the electronics used in this work is presented in Fig. \ref{fig:saSTIRAP_electronics}. The experiment was performed in a dilution refrigerator with base temperature below $\sim 20\ \textrm{mK}$. 
The sample consists of a transmon coupled to a $\lambda/4$ - coplanar waveguide (CPW) resonator, and it is made of an aluminum film with $90\ \textrm{nm}$ thickness, deposited on the surface of a silicon chip. The Josephson junctions of the transmon are realized by the shadow deposition technique, and have roughly an area of $220\times 260\ \textrm{nm}^2$. The transition frequencies of the transmon can be tuned by varying the magnetic flux threading the SQUID loop of the device, which is done by changing the current flowing through a nearby line, see Fig. \ref{fig:saSTIRAP_electronics}. The microwave pulses to control the transmon state are sent to the gate line, which is capacitively coupled to the transmon. The charging energy of the transmon $E_C \simeq h\cdot 370\ \textrm{MHz}$ and its maximum Josephson energy $E_{J\Sigma} = E_{J,1} + E_{J,2} \simeq 17.94\ \textrm{GHz}$ (with nearly identical Josephson junctions) were determined by spectroscopy measurements. The resonant frequency of the resonator, when the transmon is far detuned from it, was $f_r \simeq 5.126\ \textrm{GHz}$ with the loaded quality factor $Q \simeq 3400$. The vacuum Rabi coupling strength between the transmon and the resonator $g \simeq 41\ \textrm{MHz}$ was determined from the avoided crossing observed when the resonator and the transmon frequencies are close to each other.

Due to the coupling between the transmon and the coplanar waveguide (CPW) resonator, the state of the former is encoded into the resonant frequency of the latter, so the state of the transmon can be deduced by sending a microwave probe pulse to the resonator and measuring the reflected signal. A homodyne detection scheme was used to achieve this. The continuous microwave at frequency $f_p = 5.126\ \textrm{GHz}$, provided by a vector network analyzer (PNA-L N5230C) is split into two parts: the first of them is shaped into a probe rectangular pulse with the use of an IQ-mixer (IQ-0307LXP, m1 in Fig. \ref{fig:saSTIRAP_electronics}), while the second part serves as a LO signal for the detection part of the scheme. The reflected signal was demodulated with the same type of IQ-mixer (denoted by m2) and recorded with a high speed ADC (Acquiris U1082a).

An Agilent 33500B waveform generator was employed to provide a DC voltage to the flux bias line used to generate the  magnetic flux piercing the transmon SQUID loop. To reduce the flux noise this voltage was filtered with a passive low pass RC-filter placed at the 4K-flange of the cryostat, with cut-off frequency of  $\sim 500\ \textrm{Hz}$. A key element in the setup was the mixer denoted by m3 (see Fig. \ref{fig:saSTIRAP_electronics}), which was used to generate the microwave gate pulses for state manipulation. To do this, two channels from a Tektronix 5014B arbitrary waveform generator (AWG) were used to input modulated IF waves, and a microwave signal generator (Anritsu MG3692C) provided the local oscillator (LO) signal.
The IF waves were programmed in the waveform generator: their frequency was such that, after mixing, they would match the frequencies of the corresponding transmon transitions. 
%This also allowed us to implement the time delays and phase shifts digitally. The LO leakeage in this mixer was compensated by additional dc bias voltages applied to the ports. such that, after mixing, they would match the frequencies of the three pulses with the corresponding transmon transitions and to implement the time delays and phase shifts between the pulses. Each of the mixers had some dc bias voltages, applied to the I and Q ports, to compensate the leakage of the LO frequency. 
The LO leakeage in this mixer was compensated by additional dc bias voltages applied to the ports, and reduced down to the level of the background noise.
As a result of frequency mixing, in addition to the desired signal at the frequency  $\textrm{LO} - \textrm{IF}$, a mirror image appears at the frequency $\textrm{LO} + \textrm{IF}$, which, if left uncorrected, could produce spurious excitations to higher levels.  We employed standard single-sideband (SSB) calibration by adding the same IF waveform as that applied to the I port into the 
Q port of the IQ-mixer, but with the phase shifted by $\pi/2$. The output signal from  the mixer was filtered and then amplified in order to be able to achieve Rabi oscillations with short duration pulses.

We obtained the dependence of the transition frequencies on flux by performing two-tone spectroscopy measurements. The two-photon transition is also visible in spectroscopy at high excitation power. The parameters of the $\pi$ pulses for all three transitions (0 \--- 1, 0 \--- 2, and 1 \--- 2) were obtained from standard Rabi experiments. To find the relaxation rate $\Gamma_{01}= 0.6$ MHz we first excited the transmon with a $\pi$ pulse to the state $\ket{1}$  and then traced its decay to the ground state. Similarly,  $\Gamma_{21}= 0.83$ MHz by populating $\ket{2}$ with a two-photon $\pi$ pulse, then comparing the observed relaxation with numerical simulations.  The pure dephasing rates of the 01 and 12 transitions were determined from the Ramsey experiment to be $\Gamma_\phi^{01} \approx \Gamma_\phi^{12} \approx 0.09$ MHz. Consequently, for this particular sample the relaxation rates provide the dominant decoherence mechanism.
% making the precise determination of the additional pure dephasing noise difficult. We have verified this by performing Ramsey measurements, where two $\pi/2$ pulses with a variable time separation are applied to the system. We estimate that the pure dephasing rates \cite{jien_dephasing} are at most of the order of $0.5$ MHz.

\begin{figure}[!ht]
	\includegraphics[width=0.7\columnwidth]{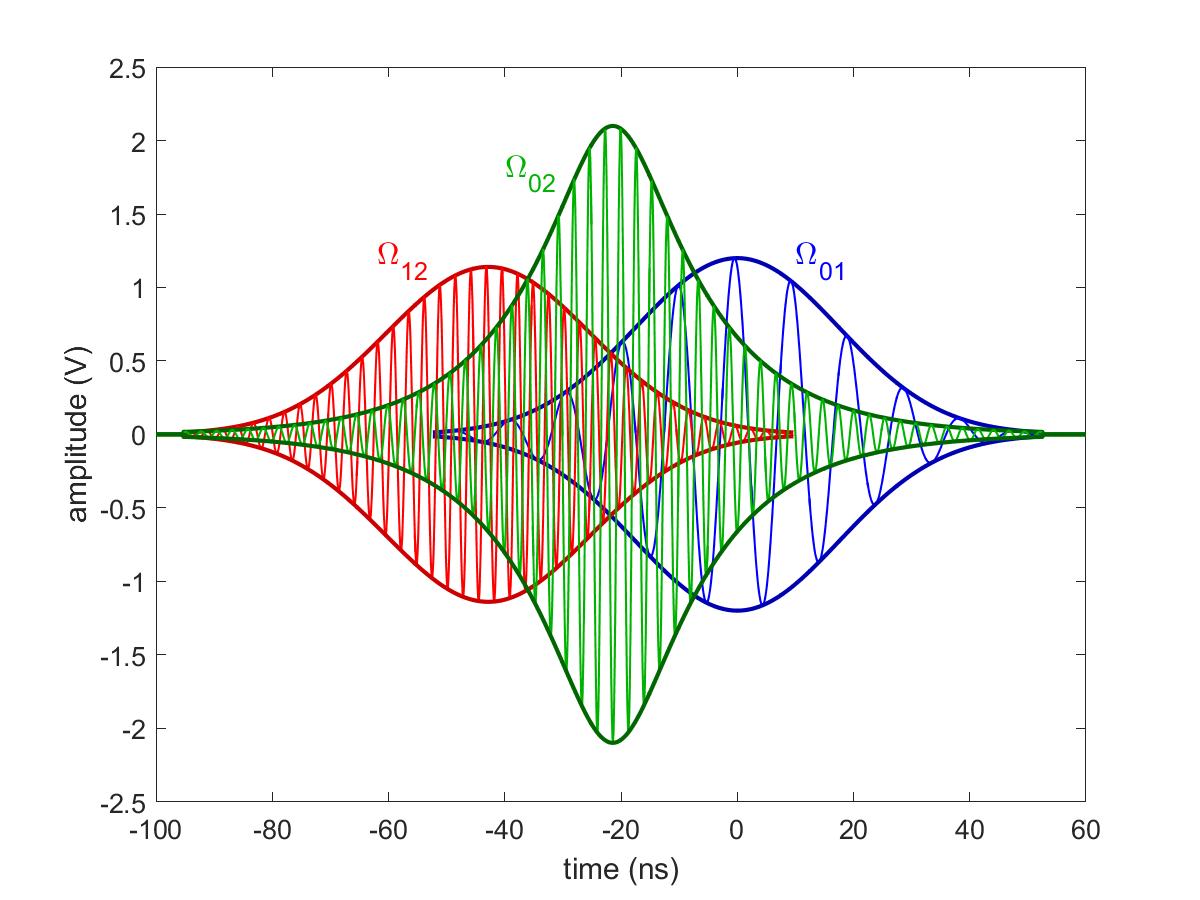}
	\caption{{\bf Pulse sequence for saSTIRAP.}
		The qutrit is controlled by three microwave pulses applied to the transitions $0-1$, $1-2$, and $0-2$. The pulses are created by an arbitrary waveform generator, leading to the corresponding Rabi couplings $\Omega_{01}$, $\Omega_{12}$, and $\Omega_{02}$.}
	\label{pulse_sequence_sa_STIRAP}
\end{figure}

Using this arrangement, we were able to generate the three overlapping gate pulses needed for the experiment.
The two STIRAP pulses were resonant with the 0 - 1 and 1- 2 transitions of the transmon, at frequencies $\omega_{01}$ and $\omega_{12}$. The third pulse was used to drive the two-photon 0 - 2 transition and had the frequency $\omega_{02}=(\omega_{01}+\omega_{12})/2$. Thus, by programming the following wavefunctions
\begin{equation}
A_{01}(t) = A_{01}\exp\left(-\frac{t^2}{2\sigma^2}\right)\times\sin(\textrm{IF}_{01}t + \phi_{01}),
\end{equation}
\begin{equation}
A_{12}(t) = A_{12}\exp\left[-\frac{(t-t_\mathrm{s})^2}{2\sigma^2}\right]\times\sin(\textrm{IF}_{12}t + \phi_{12} ),\quad \textrm{and}
\end{equation}
\begin{equation}
A_{\rm 2ph}(t) = A_{\rm 2ph}\frac{1}{\sqrt{\cosh\left[\frac{t_\mathrm{s}(t-t_\mathrm{s}/2)}{\sigma^{2}}\right]}}\times\sin(\textrm{IF}_{02}t + \phi_{\rm 2ph})
\end{equation}
into the Tektronix arbitrary waveform generator, we were able to drive the transmon with tones at $\omega_{01}$, $\omega_{12}$, $\omega_{02}$, and with the pulse envelopes required by the saSTIRAP protocol. Here the coefficients  $A_{01}$, $A_{12}$, and $A_{\rm 2ph}$ are the amplitudes of the IF pulses in volts. 

The LO frequency (common to all three pulses) was LO = $2\pi\times 6.92\ \textrm{GHz}$. The LO frequency was chosen to be sufficiently far away from all the transition frequencies to reduce even more the probability of spurious excitations. From this, we obtained the desired pulses by applying amplitude modulated waveforms to the I and Q ports of the IQ-mixer m3 with the following IF modulation frequencies:
\begin{equation}
\textrm{IF}_{01} = \textrm{LO} - \omega_{01}, \qquad \textrm{IF}_{12} = \textrm{LO} - \omega_{12}, \qquad \textrm{IF}_{02} = \textrm{LO} - \omega_{02}.
\end{equation}

The first two pulses are Gaussian shaped with standard deviation $\sigma$, and they produce the $\Omega_{01}$ and $\Omega_{12}$ drives (blue and red in Fig. \ref{pulse_sequence_sa_STIRAP}) when coupled into the corresponding transitions of the transmon. 
The third pulse $\Omega_{02}$ (green in Fig. \ref{pulse_sequence_sa_STIRAP}) has a special shape: its maximum lays in-between the $\Omega_{01}$ and $\Omega_{12}$ pulses, and it has a phase shift $\phi_{\rm 2ph}$ with respect to the Gaussian pulses. The $\Omega_{01}$ and $\Omega_{12}$ pulses were truncated at $\pm 3\sigma$ from their maxima. % so that the envelope of the pulse drops to zero at time moments which are outside of the specified intervals. 
To ensure overlap with the STIRAP pulses irrespective to the time delay $t_\mathrm{s}$, the truncation of the two-photon pulse is performed in such a way that it is zero only when both of the pulses $\Omega_{01}$ or $\Omega_{12}$ are zero (see Fig. \ref{pulse_sequence_sa_STIRAP}).

The voltages $A_{i,j}$ were converted to Rabi rates $\Omega_{i,j}$ using a calibration experiment, where the pulse amplitude is increased until a full $\pi$-pulse is created. The experiment is repeated for the 0\---1, 1\---2 and 0\---2 transitions.

An important observation is that once the phase difference between the two-photon pulse and the sum of the STIRAP phases is fixed at some reference time $t^{\rm [ref]}$, it will remain the same at any further moment in time. Suppose that the phase of the counterdiabatic two-photon pulse is $\phi_{\rm 2ph}^{\rm [ref]}$, while those of the STIRAP pulses are $\phi_{01}^{\rm [ref]}$ and $\phi_{12}^{\rm [ref]}$ at a time $t^{\rm [ref]}$; then at a later time $t$ the two-photon phase has advanced to $2 \omega_{02} (t - t^{\rm [ref]}) + 2\phi_{\rm 2ph}^{\rm [ref]}$, while the total STIRAP phase to $\omega_{01} (t - t^{\rm [ref])}) +  \phi_{01}^{\rm [ref]} + \omega_{12} (t - t^{\rm [ref]}) +  \phi_{12}^{\rm [ref]}$. The difference between these two quantities is therefore time-independent, since  $2\omega_{02} = \omega_{01} + \omega_{12}$.

The two photon driving creates small ac Stark shifts in the energy levels of the qutrit. To compensate for these, a dynamical phase correction was introduced to the drive envelopes by adding an additional phase factor $\tilde{\phi}_{ij}(t)$, as described in Methods. The corrected pulse shapes are

\begin{equation}
\tilde{A}_{01}(t) = A_{01}\exp\left(-\frac{t^2}{2\sigma^2}\right)\times\sin[\textrm{IF}_{01}t + \phi_{01} + \tilde{\phi}_{01}(t)], \label{a01}
\end{equation}
\begin{equation}
\tilde{A}_{12}(t) = A_{12}\exp\left[-\frac{(t-t_\mathrm{s})^2}{2\sigma^2}\right]\times\sin[\textrm{IF}_{12}t + \phi_{12} + \tilde{\phi}_{12}(t)],\quad \textrm{and} \label{a12}
\end{equation}
\begin{equation}
\tilde{A}_{\rm 2ph}(t) = A_{\rm 2ph}\frac{1}{\sqrt{\cosh\left[\frac{t_\mathrm{s}(t-t_\mathrm{s}/2)}{\sigma^{2}}\right]}}\times\sin[\textrm{IF}_{02}t + \phi_{\rm 2ph} + \tilde{\phi}_{02}(t)/2], \label{a02}
\end{equation}
where $\tilde{\phi}_{ij} = \int_{-\infty}^t \, \mathrm{d}t \epsilon_{ij}(t)/\hbar$ as given in the main text. Note that the phase correction coefficient $\tilde{\phi}_{02}(t)$ is divided by 2 due to the two-photon driving.

As always in experiments, the devices used for control and measurement are imperfect, and as a result deviations from the ideal case occur. Due to the finite bandwidth of the Tektronix waveform generator, the IF signals are only approximately given by the expressions Eq. \ref{a01} - \ref{a02}. Next along the control line, the single sideband cancellation procedure for the mixer m3 introduces also a phase shift $\phi^\epsilon$ in the resulting tone at the RF output. This phase shift is frequency-dependent and therefore it will be different for each of the three IF waveforms. The end result is that the gauge-invariant $\Phi$ will acquire a constant phase error $\phi_{01}^\epsilon +  \phi_{12}^\epsilon  - 2 \phi_{2\rm{ph}}^\epsilon$. This shows up as a phase shift in Fig. 5 in the main text. Next, as the signal reaches the sample, the bonding wires have a finite inductance, resulting in mismatching to the 50 Ohms line impedance and therefore in slight deformations of the pulses as seen by the transmom.

\section*{Reverse engineering of the counteradiabatic drive}

For completeness we give here a systematic derivation of the method used to find the counterdiabatic drive Hamiltonian $H_{\rm cd}$. For reference, see {\it e.g.} \cite{Berry09, Giannelli14}. Given a Hamiltonian
\begin{equation}
H_{0} (t) = \sum_{n} \lambda_{n}(t) |n(t) \rangle \langle n(t) |,
\end{equation}
where $|n(t)\rangle$ are the eigenvectors and $\lambda_{n}(t)$ are the eigenvalues, the exact solution of the Schr\"odinger equation $i\hbar \partial_{t} |\psi (t)\rangle  = H_{0}(t)|\psi (t)\rangle$ can be obtained by expanding the wavefunction into the instantaneous eigenstates $|n(t)\rangle $,
\begin{equation}
|\psi (t) \rangle = \sum_{n} c_{n} e^{i\zeta_{n} (t)}|n (t)\rangle .
\end{equation}
Using this expansion, as well as the identity
\begin{equation}
\langle n| \partial_t m\rangle = \frac{\langle n|(\partial_t H_0 )|m\rangle}{\lambda_m - \lambda_n}, \label{eq:useful_identity}
\end{equation}
obtained from taking the time derivative of $H_0 |m\rangle = \lambda_m |m\rangle$, we get
\begin{equation}
\dot{c}_n + i \dot{\zeta}_{n} c_{n} + c_{n}\langle n|\partial_t n\rangle + \frac{i}{\hbar} \lambda_{n} c_{n} + \sum_{m\neq n} c_{m} \frac{\langle n| (\partial_t H_0 ) |m\rangle}{\lambda_m - \lambda_n} e^{i (\zeta_m  - \zeta_n )} = 0 . \label{eq:full_evolution}
\end{equation}
The adiabatic theorem \cite{Born28} allows us to neglect the last term of the left hand side. Therefore in this approximation the system remains (up to a phase) in the state $|n(t)\rangle$ if it started in $|n(0)\rangle$ and there are no transitions on states $m\neq n$ during the evolution. Therefore, without loss of generality, we can take $c_{n}=1$, and obtain the equation for the phase
\begin{equation}
\dot{\zeta}_{n}(t) = - \frac{1}{\hbar} \lambda_{n} (t) + i \langle n (t)|\partial_{t} n(t)\rangle, \label{eq:zeta}
\end{equation}
which can be integrated immediately to find $\zeta (t)$ at any time $t$. Note that $\zeta (t)$ comprises a dynamical component $-\int_{0}^{t} d\tau \lambda_{n} (\tau )/\hbar $ as well as a geometric (Berry) phase part $i \int_{0}^{t} d\tau \langle n (\tau ) |\partial_{\tau }n (\tau )\rangle $.
Thus, in the adiabatic approximation the state of the system at any time $t$, starting in an
eigenstate $|n(0)\rangle$ is
\begin{equation}
|\psi (t)\rangle_{\rm ad}= e^{i\zeta_{n} (t)}|n(t) \rangle ,\label{eq:adiabatic_approx}
\end{equation}
with the phase given by the solution of \eqref{eq:zeta}.

Now, we would like to design a total Hamiltonian $H(t)$ such that the evolution of the system
under its action follows exactly the approximate adiabatic solution \eqref{eq:adiabatic_approx}.  The unitary evolution operator which would take the system along $|\psi (t)\rangle_{\rm ad}$ is
\begin{equation}
U(t) = \sum_{n} e^{i\zeta_{n} (t)} |n (t) \rangle \langle n (0)|. \label{eq:unitary}
\end{equation}
This can be implemented by reverse-engineering the evolution -- that is, given the unitary \eqref{eq:unitary} we would like to find a Hamiltonian $H(t)$ such that
\begin{equation}
i\hbar \dot{U}(t) = H(t) U(t).
\end{equation}
The formal solution for $H(t)$, from the equation above, is
\begin{equation}
H(t) = i \hbar \dot{U}(t)U^{\dag}(t).
\end{equation}
Using now Eqs. (\ref{eq:zeta}) and (\ref{eq:unitary}) we find that
\begin{equation}
H(t) = H_{0}(t) + H_{\rm cd}(t).
\end{equation}
and now we are able to identify the counteradiabatic drive Hamiltonian as
\begin{equation}
H_{\rm cd}(t) = i \hbar \sum_{n} \left[|\partial_t n(t)\rangle \langle n(t)| - \langle n(t) |\partial_t n(t)\rangle |n(t)\rangle \langle n(t)| \right] . \label{eq:counter}
\end{equation}
In the adiabatic basis, the drive Hamiltonian has only off-diagonal elements ($\langle n(t)| H_{\rm d}(t) |n(t)\rangle =0$) which can be expressed using \eqref{eq:useful_identity} as
\begin{equation}
\langle m(t)| H_{\rm cd}(t) |n(t)\rangle \vert_{n\neq m} = i \hbar \frac{\langle m(t)|(\partial_{t} H_{0})|n(t)\rangle}{\lambda_{n}(t)-\lambda_{m}(t)}.
\end{equation}

We now apply this general result to the three-level system, with energy eigenstates $|0\rangle $, $|1\rangle$, and $|2\rangle $, driven resonantly by the two fields
with frequencies
$\omega_{01}$ and $\omega_{12}$, which couple into the transitions $|0\rangle \rightarrow |1\rangle$ and $|1\rangle \rightarrow |2\rangle$
with Rabi frequencies $\Omega_{01}(t)$ (pump) respectively $\Omega_{12}(t)$ (Stokes). In the rotating wave approximation with respect to the two tones, the effective three-level Hamiltonian takes the form \cite{AT_us,ourPRB,nonabelianwallraff,stirap_ours}
% The transitions between these three states are performed by two drive signals, both having different tone. In the rotating wave approximation (RWA) the Hamiltonian of the system can be written
%\begin{equation}
%\label{eq:awg_hamiltonian}
%H_{0}^{(\delta_{01},\delta_{12})}(t) = \frac{\hbar}{2}
%\begin{bmatrix}
%0 & \Omega_{01}(t) & 0 \\
%\Omega_{01}(t) & 2\delta_{01} & \Omega_{12}(t) \\
%0 & \Omega_{12}(t) & 2(\delta_{01} + \delta_{12})
%\end{bmatrix},
%\end{equation}
%where the detunings $\delta_{01}=\omega_{01} - \omega_{01}%^{(\Omega )}$ and
%$\delta_{12} = \omega_{12} - \omega_{12}^{(\Omega )}$. 

%Since we know that STIRAP is most efficient at single-photon and two-photon resonance, we will concentrate on the case $\delta_{01}=\delta_{12}=0$. In this case (\ref{eq:awg_hamiltonian}) simplifies to
\begin{equation}
\label{eq:hamiltonian_zero_suppl}
H_{0}(t) = \frac{\hbar}{2}
\begin{bmatrix}
0 & \Omega_{01}(t) & 0 \\
\Omega_{01}(t) & 0 & \Omega_{12}(t) \\
0 & \Omega_{12}(t) & 0
\end{bmatrix}.
\end{equation}
For simplicity we have taken here the Rabi couplings $\Omega_{01}(t)$ and $\Omega_{12}(t)$ to be real. 
Using the standard parametrization of the relative strengths of the pump and Stokes frequencies by an angle $\Theta$,
\begin{equation}
\tan \Theta (t) = \frac{\Omega_{01}(t)}{\Omega_{12}(t)}, \label{eq:thetadef}
\end{equation}
we can express the eigenvectors of this Hamiltonian as
\begin{equation}
\label{eq:eigenvectors}
\begin{aligned}
\ket{n_{+}(t)} &= \frac{1}{\sqrt{2}}\sin{\Theta (t)}\ket{0}+ \frac{1}{\sqrt{2}}\ket{1} + \frac{1}{\sqrt{2}}\cos{\Theta (t)}\ket{2},\\
\ket{n_{-}(t)} &= \frac{1}{\sqrt{2}}\sin{\Theta (t)}\ket{0}- \frac{1}{\sqrt{2}}\ket{1} + \frac{1}{\sqrt{2}}\cos{\Theta (t)}\ket{2},\\
\ket{n_{0}(t)} &= \cos{\Theta (t)}\ket{0} - \sin{\Theta (t)}\ket{2},
\end{aligned}
\end{equation}
and
corresponding eigenvalues
\begin{equation}
\label{eq:eigenvalues}
\begin{aligned}
\lambda_{+}(t) &= \frac{\hbar}{2} \sqrt{\Omega_{01}^{2}(t) + \Omega_{12}^{2}(t)},\\
\lambda_{-}(t) &= -\frac{\hbar}{2} \sqrt{\Omega_{01}^{2}(t) + \Omega_{12}^{2}(t)},\\
\lambda_{0}(t) &= 0.
\end{aligned}
\end{equation}
The last eigenvector is a dark state.

Now, to identify the driving Hamiltonian, we plug in the expression for eigenvectors \eqref{eq:eigenvectors} in \eqref{eq:counter}. After some algebraic calculations and noticing that the terms $\langle n(t)|\partial_t n(t)\rangle$ are all zero, we obtain a remarkably simple result,
\begin{equation}
\label{eq:finalcounterdrive_suppl}
H_{\rm cd}(t) = \frac{\hbar}{2}
\begin{bmatrix}
0 & 0 & \Omega_{02}(t) e^{i \pi/2}\\
0 & 0 & 0 \\
\Omega_{02}(t) e^{-i \pi/2}& 0 & 0
\end{bmatrix},
\end{equation}
where $\Omega_{02}(t)=2\dot{\Theta}(t)$.
Integrating the absolute value of $\Omega_{\rm cd}(t)$ between an initial time $t_{\rm in}$ and a final time $t_{\rm fin}$ we get the pulse area condition
\begin{equation}
\int_{t_{\rm in}}^{t_{\rm fin}} |\Omega_{\rm cd} (t')| dt' = 2 \Theta (t_{\rm fin}) - 2 \Theta (t_{\rm in}).
\end{equation}
Thus, if the drive Hamiltonian is intended to correct a full STIRAP, for which $\Theta (t_{\rm in})= 0$ and $\Theta (t_{\rm fin})= \pi /2$,  a total area of $\pi$ is needed for the counterdrive pulse. Note also that the quantum control implemented by \eqref{eq:finalcounterdrive_suppl} -- or in general by \eqref{eq:counter} --  implements a standard linear evolution and it is also state-independent. %The schematic representation of the action of the counterdiabatic pulse on the Majorana sphere is shown in \ref{Majorana}.

% \begin{figure}[!ht]
% 	\includegraphics[width=0.7\columnwidth]{./Figures_saSTIRAP/Majorana_representation.pdf}
% 	\caption{{\bf saSTIRAP in the Majorana stellar representation.}
% 		Majorana stellar representation of the process, where the state of the qutrit is characterized by the two roots (red and yellow colours) of the Majorana polynomial, presented in spherical coordinates.
% 		The counterdiabatic pulse corrects the trajectories of these roots (dashed yellow and red lines) and brings them to the two meridians at the intersection of the $xOz$ plane with the sphere.}
% 	\label{Majorana}
% \end{figure}

In the experiment the pump and Stokes pulses of the STIRAP protocol are Gaussians of equal width $\sigma$ delayed with respect to each other by a separation time $t_{\rm s}$,
\begin{eqnarray}
\label{eq:gaussian_envelopes_S}
%\begin{aligned}
%\Omega_{01}(t) = \Omega\exp\left[-\frac{(t - t_{s}/2)^2}{2\sigma^2}\right], \\
%\Omega_{12}(t) = \Omega\exp\left[-\frac{(t + t_{s}/2)^2}{2\sigma^2}\right], \\
\Omega_{01}(t) &=& \Omega_{01}\exp\left[-\frac{t^2}{2\sigma^2}\right], \\
\Omega_{12}(t) &=& \Omega_{12}\exp\left[-\frac{(t - t_{s})^2}{2\sigma^2}\right],
%\end{aligned}
\end{eqnarray}
where time is measured from the maximum of the $0 - 1$ pulse. In this convention, STIRAP is realized at negative times $t_{\rm s}$, while the intuitive sequence corresponds to positive $t_{\rm s}$.

From \eqref{eq:thetadef} we obtain
\begin{equation}
\dot{\Theta}(t) = \frac{\dot{\Omega}_{01}(t)\Omega_{12}(t) - \Omega_{01}(t)\dot{\Omega}_{12}(t)}{\Omega_{01}^{2}(t) + \Omega_{12}^{2}(t)}.
\end{equation}
In the case $\Omega_{01}=\Omega_{12}$ we get
\begin{equation}
\dot{\Theta} (t)= - \frac{t_{\rm s}}{2 \sigma^2} \sin \left[2 \Theta (t)\right],
\end{equation}
or
\begin{equation}
\dot{\Theta} = -\frac{t_{\rm s}}{2\sigma^2}\frac{1}{\cosh\left[ \frac{t_{\rm s}}{\sigma^2}\left(t-\frac{t_{\rm s}}{2}\right)\right]}.
\label{eq:correctionform}
\end{equation}

Furthermore, in this simple all-resonant configuration it is straigthforward to derive the local adiabaticity condition for the STIRAP pulses (valid for any pulse shape), which reads
\begin{equation}
|\dot{\Theta}(t)| \ll \sqrt{\Omega_{01}^2(t) + \Omega_{12}^2(t)}. \label{eq:adiaatic_condition}
\end{equation}
Retaining the nonadiabatic terms in \eqref{eq:full_evolution}, with the solution \eqref{eq:zeta} and the substitution $c_{\rm \pm}(t) = \tilde{c}_{\pm}(t) \exp \left[-\frac{1}{\hbar} \int_{0}^{t} dt' \lambda_{\pm}(t')\right]$ (thus separating the trivial dynamical evolution from the adiabatic one), we obtain
\begin{equation}
\begin{aligned}
\dot{c}_{0}(t) &= -\frac{\dot{\Theta}(t)}{\sqrt{2}}\tilde{c}_{+}(t) - \frac{\dot{\Theta}(t)}{\sqrt{2}}\tilde{c}_{-}(t),\\
\dot{\tilde{c}}_{+}(t) &= -\frac{i}{2}\sqrt{\Omega_{01}^{2}(t)+\Omega_{12}^2(t)}\tilde{c}_{+}(t) - \frac{\dot{\Theta}(t)}{\sqrt{2}}c_{0}(t),\\
\dot{\tilde{c}}_{-}(t) &= \frac{i}{2}\sqrt{\Omega_{01}^{2}(t)+\Omega_{12}^2(t)}\tilde{c}_{+}(t) - \frac{\dot{\Theta}(t)}{\sqrt{2}}c_{0}(t).
\end{aligned}
\end{equation}
Thus the condition \eqref{eq:adiaatic_condition} ensures that there are no transitions between the instantaneous adiabatic states during the evolution.

\section*{Synthetic Peierls couplings on the triangular plaquette}

The loop driving configuration used in the experiment produces a nontrivial gauge structure, as shown in the schematic of Fig. 1 d) in the main text. In this picture, the internal states of the transmon can be regarded as sites in a synthetic dimension \cite{Lewenstein2014}.

To explore this concept in detail, we adapt some of the results obtained earlier in \cite{spatial} to our setup. Consider what happens under local U(1) transformations with phases $\chi_{0}$, $\chi_{1}$, and $\chi_{2}$ on each of the three states. The unitary that achieves this is
%\begin{equation}
%U =  \left(\begin{array}{ccc} e^{-i \chi_{0}}  & 0 & 0 \\ 0 & e^{-i \chi_{1}} & 0 \\ 0 & 0 & e^{-i \chi_{2}} \end{array} \right),
%\end{equation}
\begin{equation}
U =  e^{-i \chi_{0}}  |0\rangle\langle 0| + e^{-i \chi_{1}} |1\rangle \langle 1| + e^{-i \chi_{2}} |2\rangle \langle 2|,
\end{equation}
leading to a transformed Hamiltonian $H' = U H U^{\dag}$ and
%\begin{equation}
%H' = \frac{\hbar}{2} \left(\begin{array}{ccc} 0 & \Omega_{01} e^{i \phi_{01}'} & \Omega_{02} e^{i \phi_{02}'} \\ \Omega_{01} e^{-i \phi_{01}'} & 0 & \Omega_{12} e^{i %\phi_{12}'} \\
%\Omega_{02} e^{-i \phi_{02}'} & \Omega_{12} e^{-i \phi_{12}'} & 0 \end{array} \right). \label{nurmi}
%\end{equation}
\begin{equation}
H' = \frac{\hbar}{2} \Omega_{01} e^{i \phi_{01}'} |0\rangle \langle 1| + \frac{\hbar}{2} \Omega_{12} e^{i \phi_{12}'} |1\rangle \langle 2|
+ \frac{\hbar}{2}\Omega_{02} e^{i \phi_{02}'} |0\rangle \langle 2| + \mathrm{h.c.} . \label{nurmi}
\end{equation}
The equations for the phases $\chi_{0}$, $\chi_{1}$, and $\chi_{2}$ are
\begin{equation}
\left[\begin{array}{ccc} -1 & 1 & 0 \\ 0 & -1 & 1 \\ 1 & 0 & -1 \end{array} \right] \left[\begin{array}{c} \chi_0 \\ \chi_1 \\ \chi_2 \end{array} \right] = \left[\begin{array}{c} \phi_{01}'-\phi_{01} \\ \phi_{12}'-\phi_{12} \\  \phi_{20}'-\phi_{20} \end{array} \right]. \label{matrix_form_gauge}
\end{equation}
Here, to put in evidence the circularity under the permutation of indices 0, 1, and 2, we have replaced $\phi_{02} = - \phi_{20}$, and $\phi_{02}' = - \phi_{20}'$. A first observation is that the $3\times 3$ matrix appearing in the equation above is singular. Given a choice of desired phases $\phi_{01}'$, $\phi_{12}'$, and $\phi_{20}'$, because the matrix is not invertible, we are not guaranteed that we can find the local gauge phases $\chi_{0}$, $\chi_{1}$, and $\chi_{2}$ such that the form $H'$ is obtained.
Indeed, by adding up the three equations from the matrix form \eqref{matrix_form_gauge} we obtain the constraint that the sum of the phases $\phi_{ij}$ (in the circular order for the indices $i,j$) before and after the transformation must be the same
\begin{equation}
\Phi = \phi_{01}' + \phi_{12}' + \phi_{20}' = \phi_{01} + \phi_{12} + \phi_{20}.
\end{equation}
This constraint implies that we cannot eliminate all three phases simultaneously, and it results from the existence, in the loop driving configuration, of three nonzero Rabi frequencies $\Omega_{01}$, $\Omega_{12}$, and $\Omega_{02}$. If any one of these Rabi frequencies would be zero, the phase elimination could be done. Suppose for example that $\Omega_{02}=0$. In this case we can obtain the Hamiltonian \eqref{nurmi} with all the elements real,
by choosing for example $\chi_{0}=0$, $\chi_{1}= - \phi_{01}$, and $\chi_{2} = -\phi_{01} - \phi_{12}$. However, if $\Omega_{02} \neq 0$, then the same choice of gauges leads to 
$\phi_{01}' = \phi_{12}'=0$ and $\phi_{20} = \Phi$. 
From an experimental point of view the existence of this constraint provides us with a useful ``knob'' for controlling the system, through adjustments of the phase $\Phi$. By arranging $\chi_{0},\chi_{1},\chi_{2}$ such that $\phi_{01}' = \phi_{12}'=0$ we get a convenient final form for the Hamiltonian
\begin{equation}
H = \frac{\hbar}{2} |0\rangle \langle 1|  + \frac{\hbar}{2} \Omega_{12} |1\rangle \langle 2| +
\frac{\hbar}{2} \Omega_{02} e^{-i \Phi}|0\rangle\langle 2| + \mathrm{h.c.}.\label{eq:full_hamiltonian_standard_suppl}
\end{equation}
In this gauge, the STIRAP part is $H_0 = (\hbar /2) \Omega_{01} |0\rangle \langle 1| + (\hbar/2) \Omega_{12} |1\rangle \langle 2| + \mathrm{h.c.}$, while the counterdiabatic part needed for saSTIRAP is realized at $\Phi = -\pi/2$ and it takes the familiar form  $H_{\rm cd} = i(\hbar /2) \Omega_{02}(t) |0\rangle \langle 2| + \mathrm{h.c.}$ \cite{Giannelli14}.  Seen as an Aharonov-Bohm effect, this situation corresponds to constructive interference on the state $|2\rangle$ between the direct two-photon drive and the STIRAP process. 

%Also note that the  Hamiltonian (\ref{eq:full_hamiltonian_standard_suppl}) with $\Phi = -\pi/2$ models a spin-1 particle in a magnetic field with components $(\Omega_{01}, \Omega_{12}, -\Omega_{02})$. Here the Pauli-matrices $\sigma_{01}^{x}$, $\sigma_{12}^{x}$, and $\sigma_{02}^{y}$ are representations of the spin-1 angular momentum operators since they satisfy the angular momentum commutation relations $[\sigma_{01}^{x}, \sigma_{12}^{x}] = i \sigma_{02}^{y}$ and circular permutations thereof.

Finally, one notices that \eqref{matrix_form_gauge}, which defines the change in the gauge, is the discrete-space analogous of the well-known relation $\vec{A}' = \vec{A} + \nabla \chi$, which describes a change of gauge for the vector magnetic field, written for a lattice with 3 positions corresponding to the ``localized'' states $|0\rangle$, $|1\rangle$, and $|2\rangle$. The differences $\chi_1 - \chi_0$, $\chi_2 - \chi_1$, and $\chi_0 - \chi_2$ which appear in \eqref{matrix_form_gauge} are the discrete version of the continuous space derivative $\nabla \chi$. The sum of the phases $\Phi$ is the discrete version of the line integral $\oint \vec{A} {\rm d}\vec{l}$, which is the magnetic flux penetrating the plaquette formed by the three sites.

To make this more precise, let us define a lattice gauge Hamiltonian model: a spinless particle that can jump between three sites, 0,1, and 2. In the second order quantization, the states with one particle at the site 0, 1, or 2 can be written respectively as $|1,0,0\rangle $, $|0,1,0\rangle $, and $|0,0,1\rangle $. Formally, we can identify these states with the standard qutrit basis $|0\rangle , |1\rangle , |2\rangle$,
\begin{eqnarray}
|1,0,0\rangle \equiv |0\rangle , |0,1,0\rangle \equiv |1\rangle , |0,0,1\rangle \equiv |2\rangle .
\end{eqnarray}
Then the Hamiltonian takes the form
\begin{equation}
H = \frac{\hbar}{2} \sum_{<j,k>} \Omega_{jk} e^{i \phi_{jk}} c_{j}^{\dag} c_{k},
\end{equation}
with the convention $\phi_{jk} = -\phi_{kj}$.
As usual in lattice field theory, the complex factors
\begin{equation}
U_{jk} = U_{kj}^{*} = e^{i \phi_{jk}} \in U(1),
\end{equation}
are referred to as link variables (Peierls phases). Then the Aharonov-Bohm phase accumulated due to transport across the plaquette (loop) is
\begin{equation}
e^{i \Phi } = \prod_{\bigtriangleup} U_{jk} = U_{01}U_{12}U_{20} = e^{i (\phi_{01} + \phi_{12} + \phi_{20} )},
\end{equation}
where the product is taken in circular order.
Now, we can introduce a gauge potential ${\bf \cal A}$ in the standard way, such that the integral of this quantity between two points $j$ and $k$ equals the phase $\phi_{jk}$,
\begin{equation}
\phi_{jk} = \frac{1}{\hbar} \int_{j}^{k} {\bf \cal A} d{\bf l}.
\end{equation}
Thus, we consider that the gauge field ${\bf \cal A}$ is defined along the links that connect the sites.
If this gauge field is produced by a standard magnetic field piercing the plaquette, then ${\bf \cal A} = q A$, where $q$ is the charge, and $\frac{1}{\hbar} \int_{j}^{k} {\bf \cal A} d{\bf l}$
is referred to as magnetic phase factor.

Then, we define a gauge transformation  ${\mathbf \mathcal A} \rightarrow {\mathbf \mathcal A}' = {\mathbf \mathcal A} + {\mathbf \nabla} \chi$. This changes the link variables as
\begin{equation}
U_{jk} \rightarrow U_{jk}' = e^{\frac{i}{\hbar} (\chi_{k}-\chi_{j})} U_{jk},
\end{equation}
resulting for the entire loop in
\begin{equation}
e^{i \Phi'} = \prod_{\bigtriangleup} U_{jk}' =  e^{\sum_{\bigtriangleup} \frac{i}{\hbar} (\chi_{k}-\chi_{j})} \prod_{\bigtriangleup} U_{jk} = \prod_{\bigtriangleup} U_{jk} = e^{i \Phi}.
\end{equation}
Here the indices are summed in circular order.  Thus, $\Phi$ is gauge invariant, and via Stoke's theorem $\oint {\mathbf \mathcal A} d {\bf l} = \int\int {\mathbf \mathcal B} d{\bf s}$. If the gauge field is the magnetic vector potential, ${\mathbf \mathcal A} = q {\bf A}$, then ${\mathbf B} = \nabla \times A$ is the magnetic field  and
\begin{equation}
\Phi = 2 \pi \frac{q}{h} \int\int {\bf B} d{\bf s}
\end{equation}
is $2\pi$ times the magnetic flux expressed in units of flux quanta $h/q$.

% For your review copy (i.e., the file you initially send in for
% evaluation), you can use the {figure} environment and the
% \includegraphics command to stream your figures into the text, placing
% all figures at the end.  For the final, revised manuscript for
% acceptance and production, however, PostScript or other graphics
% should not be streamed into your compliled file.  Instead, set
% captions as simple paragraphs (with a \noindent tag), setting them
% off from the rest of the text with a \clearpage as shown  below, and
% submit figures as separate files according to the Art Department's
% instructions.

%% \clearpage

%% \noindent {\bf Fig. 1.} Please do not use figure environments to set
%% up your figures in the final (post-peer-review) draft, do not include graphics in your
%% source code, and do not cite figures in the text using \LaTeX\
%% \verb+\ref+ commands.  Instead, simply refer to the figure numbers in
%% the text per {\it Science\/} style, and include the list of captions at
%% the end of the document, coded as ordinary paragraphs as shown in the
%% \texttt{scifile.tex} template file.  Your actual figure files should
%% be submitted separately.

\end{document}